\documentclass[12pt]{article}

\usepackage{jheppub}

\usepackage[T1]{fontenc}

\usepackage{graphicx}
\usepackage{amsfonts}
\usepackage{float}

\newcommand{\beq}{\begin{equation}}
\newcommand{\eeq}{\end{equation}}
\newcommand{\bdm}{\begin{displaymath}}
\newcommand{\edm}{\end{displaymath}}
\newcommand{\beqr}{\begin{eqnarray}}
\newcommand{\eeqr}{\end{eqnarray}}
\newcommand{\beqrn}{\begin{eqnarray*}}
\newcommand{\eeqrn}{\end{eqnarray*}}

\title{Two Species of Vortices in  Massive Gauged Non-linear Sigma Models}

\author[a]{A. Alonso-Izquierdo\note{Corresponding author.},}
\author[b]{W. Garc\'{\i}a Fuertes,}
\author[c]{J. Mateos Guilarte}

\affiliation[a]{Departamento de Matem\'atica Aplicada, Facultad de Ciencias Agrarias y Ambientales, \\ Universidad de Salamanca, E-37008 Salamanca, Spain.}
\affiliation[b]{Departamento de F\'{\i}sica, Facultad de Ciencias,\\  Universidad de Oviedo,
E-33007 Oviedo, Spain.}
\affiliation[c]{Departamento de F\'{\i}sica Fundamental, Facultad de Ciencias,\\  Universidad de Salamanca,
E-37008 Salamanca, Spain.}

\emailAdd{alonsoiz@usal.es}
\emailAdd{wifredo@uniovi.es}
\emailAdd{guilarte@usal.es}

\abstract{Non-linear sigma models with scalar fields taking values on $\mathbb{C}\mathbb{P}^n$ complex manifolds are addressed. In the simplest $n=1$ case, where the target manifold is the $\mathbb{S}^2$ sphere, we describe the scalar fields by means of stereographic maps. In this case when the $\mathbb{U}(1)$ symmetry is gauged and Maxwell and mass terms are allowed, the model accommodates stable self-dual vortices of two kinds with different energies per unit length and where the Higgs field winds at the cores around the two opposite poles of the sphere. Allowing for dielectric functions in the magnetic field, similar and richer self-dual vortices of different species in the south and north charts can be found by slightly modifying the potential. Two different situations are envisaged: either the vacuum orbit lies on a parallel in the sphere, or one pole and the same parallel form the vacuum orbit. Besides the self-dual vortices of two species, there exist BPS domain walls in the second case. Replacing the Maxwell contribution of the gauge field to the action by the second Chern-Simons secondary class, only possible in $(2+1)$-dimensional Minkowski space-time, new BPS topological defects of two species appear. Namely, both BPS vortices and domain ribbons in the south and the north charts exist because the vacuum orbit consits of the two poles and one parallel. Formulation of the gauged $\mathbb{C}\mathbb{P}^2$ model in a Reference chart shows a self-dual structure such that BPS semi-local vortices exist. The transition functions to the second or third charts break the $\mathbb{U}(1)\times\mathbb{S}\mathbb{U}(2)$ semi-local symmetry, but there is still room for standard self-dual vortices of the second species. The same structures encompassing $N$ complex scalar fields are easily generalized to gauged $\mathbb{C}\mathbb{P}^N$ models.}


\begin{document}
\maketitle
\flushbottom

\section{Introduction}

Bogomolny-Prasad-Sommerfield topological defects in gauge theories polluted with scalar fields in the Higgs phase are solutions of first-order ODE systems with specific asymptotic behaviour. Besides of forming lumps of energy, or energy per $(length)^d$  these self-dual solitons are stable and were characterized in the seminal papers \cite{prasad,bogom}. Their important r$\hat{\rm o}$le in supersymmetric gauge theories as states giving rise to topological central charges of the SUSY algebra was recognized in \cite{Olive}. Self-dual vortices, both of the Nielsen-Olesen \cite{nieole} and the Chern-Simons \cite{Jackiw} type, are among the most interesting BPS defects and have been much studied in purely bosonic as well as in supersymmetric theories, see \cite{susyvort1,susyvort2,susyvort3} for instance. The existence of BPS equations and a BPS bound in a non-linear sigma model with the target manifold an $\mathbb{S}^2$-sphere and $\mathbb{R}^{1,2}$ as Minkowski space-time was investigated by B.J. Schroers in the short letter \cite{Schroer}. In that work, the author introduced a $\mathbb{U}(1)$ gauge field minimally coupled to the scalar fields via covariant derivatives, whereas a Maxwell term allowed for the presence of planar vector bosons. This approach led to BPS solitons, which were found to be akin to baby skyrmyons or $\mathbb{C}\mathbb{P}^1$-lumps existing in other $(2+1)$-dimensional models. There have also been other approaches to the problem, as the paper \cite{Nard} were the non-linear sigma model with non-abelian $O(3)$ gauge symmetry and Chern-Simons interaction was considered or the general study \cite{ bapt} of abelian vortex equations where both the base space and the target are K\"{a}hler manifolds. Nevertheless, one may think that in a system of this kind, suitable for describing the dynamics of ideal charged bosonic plasmas, topological defects of the Nielsen-Olesen vortex type should exist. In the guise of two-dimensional instanton self-dual vortices were discovered by Nitta and Vinci in the $(1+1)$-dimensional $\mathbb{C}\mathbb{P}^1$-sigma model with extended ${\cal N}=(2,2)$ supersymmetry,
see Reference \cite{nivi}. Closely related fractional instantons where  twisted boundary conditions on the worldsheet replace the potential term have been recently described in \cite{nitta14}, but in this case the instantons combine in stable bions; they are relevant to confinement and the the so-called resurgence phenomenon of the Borel resummed perturbative series, where renormalon singularities are cancelled by bions, see \cite{minisa} and references therein. Here we shall show that in a slight modification of the Schroers scenario, self-dual topological vortices do indeed exist and enjoy quite standard properties, except that there are two species of magnetic flux tubes, i.e., we shall describe in detail the Nitta-Vinci vortices in a purely bosonic model in $(3+1)$-dimensions.

One-dimensional topological defects of the kink type have been unveiled in massive non-linear $\mathbb{S}^2$-sigma models in $(1+1)$-dimensions. In that case the search was made possible because of the Hamilton-Jacobi separability in elliptic coordinates of the Neumann problem, a solvable dynamical system which is tantamount to the search for kinks in the non-linear $\mathbb{S}^2$-sigma model with a non-degenerate mass spectrum, see \cite{amaj1,amaj2}. The procedure also worked for finding kinks in a hybrid non-linear $\mathbb{S}^2$-sigma/Ginzburg-Landau 2D model \cite{amaj3}. As in this latter case, we shall see that only a very precise choice of the potential is compatible with the existence of self-dual vortices in the massive gauged non-linear $\mathbb{S}^2$-sigma model. To build the $\mathbb{U}(1)$ gauge theory out of the $\mathbb{O}(3)$ background symmetry of the $\mathbb{S}^2$-sphere, our proposal is to gauge the stereographic coordinates rather than the original fields. By doing so, the potential can be chosen in such a way that self-duality is guaranteed simultaneously in both the south- and north- charts of the sphere. Although the vorticial solutions corresponding to both charts have different energies per unit  length, there is a local version of the Bogomolny bound that ensures their separate stability.

The BPS structure of this model is compatible with the modification of the magnetic field by a dielectric function. By means of this generalization
and the choice of a dielectric factor that is well behaved in the two charts of the sphere, we find families of self-dual vortices carrying scalar profiles and magnetic fields that are susceptible to being modified almost at will. These new vortices belong to the class
discovered in \cite{leenam} but also appear in two species attached respectively to the north and south poles. It is interesting to point out that
the Nitta-Vinci vortices correspond to the choice of a constant dielectric function. In particular, we shall show that self-dual vortices of two species exist which are akin to those unveiled in \cite{Fuerguil} in the commutative limit but choosing $\mathbb{S}^2$ as the target manifold.
We also remark that the dielectric function $H(\vert\phi\vert)$ may be chosen such that the scalar field reach its vacuum value both at one parallel and either the south or the north pole.
In these cases the model not only support self-dual vortices of two species but also BPS domain walls with finite energy per unit surface living either in the south or the north charts. The domain walls grow from one-dimensional kinks akin to those unveiled in \cite{Nitta, Nitta1, Nitta2}. Things are more interesting replacing Maxwell by Chern-Simons dynamics on the gauge field. The system is forced to be defined in a $(2+1)$-dimensional Minkowski
space-time but the solution of the Gauss law for static fields produces a energy with a fixed dielectric function. Then, one chooses the potential such that self-duality equations arise in both charts and discover two species of self dual vortices  analogue to those existing in the Abelian Chern-Simons-Higgs planar gauge theory, see \cite{Jackiw}. We mention that replacement of $\mathbb{R}^2$ by $\mathbb{S}^2$
as the scalar field target space is mandatory in our model. Moreover, the Chern-Simons dielectric function prompts a vacuum orbit formed by the two poles and one parallel. Thus, not only two species of self-dual vortices exist but also BPS ribbons with finite energy per unit length emerge
in two types living each species in one chart.

The $\mathbb{S}^2$ sphere as a complex manifold is the $\mathbb{C}\mathbb{P}^1$ compactification of $\mathbb{C}$, and the stereographic version of the round metric, which is the key ingredient to giving the right properties to the vortices on the sphere, becomes precisely the Fubini-Study metric of that K\"{a}hler manifold. Thus, it is natural to think that some kind of well-behaved vortices, sharing many features with $\mathbb{S}^2$ vortices, should also exist in the higher rank non-linear $\mathbb{C}\mathbb{P}^n$-sigma Abelian Higgs model. In the last section of the paper we shall show that this is in fact the case. There is, however, an important novelty: in one of the $\mathbb{C}\mathbb{P}^n$-charts there exist BPS semi-local topological solitons, with or without vorticity, which are the cousins of the semi-local defects discovered in the scalar sector of Electroweak Theory when the weak angle is $\frac{\pi}{2}$ and described in References \cite{vaachu,hind,gors}. In the $n-1$ remaining charts, only purely vorticial self-dual vortices exist, all of them of the second species.

\section{Construction of the non-linear $S^2$-sigma model}

We begin with a system of three scalar fields $\Phi_a$, $a=1,2,3$, taking values on a $\mathbb{S}^2$-sphere: $\Phi_a(x^\mu): \mathbb{R}^{1,3} \rightarrow  \mathbb{S}^2$ where $\Phi_1^2(x^\mu)+\Phi_2^2(x^\mu)+\Phi_3^2(x^\mu)=\rho^2$ and $x^\mu\equiv (x^0,x^1,x^2,x^3)\in \mathbb{R}^{1,3}$. We denote $x^\mu\cdot x_\mu=g_{\mu\nu}x^\mu x^\nu$ with $g_{\mu\nu}={\rm diag}(1,-1,-1,-1)$. Stereographic projections from either the south or north poles of the $\mathbb{S}^2$-sphere to the $\mathbb{R}^2$ plane provide a minimal atlas formed by the south and north charts. The respective stereographic coordinates for the south and north charts are the complex scalar fields
\bdm
\phi(x^\mu)=\rho\frac{\Phi_1(x^\mu)+i \Phi_2(x^\mu)}{\rho-\Phi_3(x^\mu)}\hspace{2cm}\psi(x^\mu)=\rho\frac{\Phi_1(x^\mu)-i \Phi_2(x^\mu)}{\rho+\Phi_3(x^\mu)} \, \ ,
\edm
whereas the inverse mapping, e.g., from the south chart, reads:
\begin{equation}
\Phi_1(x^\mu)+i\Phi_2(x^\mu)=\frac{2\rho^2\phi(x^\mu)}{\rho^2+| \phi(x^\mu)|^2} \quad , \quad \Phi_3(x^\mu)=\rho\frac{| \phi(x^\mu)|^2-\rho^2}{\rho^2+| \phi(x^\mu)|^2}\label{invstp}\quad .
\end{equation}
The transformation $\phi=\frac{\rho^2}{\psi^*}$ from the south to the north chart
reverses the orientation. The reason for choosing this option is to deal with scalar fields coupled to the gauge field with identical electric charges in both charts. The massless Lagrangian ${\cal L}_\Phi$ describing the dynamics of the $\Phi$ fields, in terms of the south-chart field, becomes:
\bdm
{\cal L}_\Phi=\frac{1}{2}\frac{4\rho^4}{(\rho^2+|\phi|^2)^2}\, \partial_\mu \phi^* \partial^\mu \phi.
\edm
The global $U(1)$ symmetry is made local following the standard procedure: a gauge field $A_\mu$ enters the system and supplements the local $U(1)$ transformation $\phi\rightarrow e^{i e \chi(x)}\phi$ with the gauge transformation $A_\mu\rightarrow A_\mu+\partial_\mu\chi$. A potential energy density yielding spontaneous symmetry breaking and the Higgs mechanism can also be introduced. All this leads to the Lagrangian of a gauged  massive Abelian non-linear $\mathbb{S}^2$-sigma model of the form
\beq
{\cal L}_S=-\frac{1}{4} F_{\mu\nu} F^{\mu\nu} +\frac{1}{2}\frac{4\rho^4}{(\rho^2+|\phi|^2)^2} D_\mu \phi^* D^\mu \phi-U_S(|\phi|^2)\label{lagphi}\, \, .
\eeq
The covariant derivative and the electromagnetic tensor are defined in the usual way: $D_\mu\phi=\partial_\mu\phi-i e A_\mu\phi$, $F_{\mu\nu}=\partial_\mu A_\nu-\partial_\nu A_\mu$. Under the change of coordinates $\phi=\frac{\rho^2}{\psi^*}$, the covariant derivatives and the potential energy density in the north chart become:
\beq
D_\mu\phi=-\frac{\rho^2}{(\psi^*)^2}D_\mu\psi^* \, \, \, , \quad U_N(|\psi|^2)=U_S\Big(\frac{\rho^4}{|\psi|^2}\Big)\label{cambioderi}
\eeq
in such a way that the Lagrangian in this chart now reads:
\beq
{\cal L}_N=-\frac{1}{4} F_{\mu\nu} F^{\mu\nu} +\frac{1}{2}\frac{4\rho^4}{(\rho^2+|\psi|^2)^2} D_\mu \psi^* D^\mu \psi-U_N(|\psi|^2)\label{lagpsi}.
\eeq

\subsection{Bogomolny splitting and self-dual vortices}

Lagrangians such as (\ref{lagphi}) and (\ref{lagpsi}) with a function of the scalar field multiplying the covariant derivative term were studied by M. A. Lohe in \cite{lohe}. Other references in which  models with a similar structure were investigated are, for instance, \cite{juwi,bhsm,bchm}. In a field theory encompassing scalar and gauge fields where the Lagrangian density is of the general form
\bdm
{\cal L}=-\frac{1}{4} F_{\mu\nu} F^{\mu\nu} +\frac{1}{2} g(|\varphi|^2) D_\mu \varphi^* D^\mu \varphi-\frac{1}{2} W^2(|\varphi|^2),
\edm
and both the function $g(|\varphi|^2)$ and the potential energy density are semi-definite positive, it is possible to write the energy per unit length of static and $x_3\in [-L/2,L/2]$-independent configurations \`{a} la Bogomolny
\bdm
E/L=\int d^2 x\left\{\frac{1}{2} \left(B\pm W(|\varphi|^2)\right)^2+\frac{1}{2}g(|\varphi|^2) |D_1\varphi\mp i D_2\varphi|^2\mp \left(W(|\varphi|^2) B+ R\right)\right\}
\edm
after taking the Weyl plus axial gauge: $A_0=A_3=0$. Here, $B=F_{12}$ is the magnetic field and $R=-\frac{i}{2} g(|\varphi|^2) \varepsilon_{ij} D_i\varphi^* D_j \varphi, i,j=1,2$ can be rewritten as
\bdm
R=-\frac{i}{2} \varepsilon_{ij} \partial_i [F(|\varphi|^2) (\partial_j \ln \varphi-ie A_j)]+ \frac{1}{2}e F(|\varphi|^2) B,
\edm
where $F$ is a primitive of $g$ such that $F(0)=0$ in order to avoid singularities in the logarithm. Choosing a potential energy density
\beq
U(\varphi)=\frac{1}{2} W^2(|\varphi|^2)=\frac{e^2}{8}(F(|\varphi|^2)-a^2)^2,\label{sdpot}
\eeq
where $a^2$ belongs to the range $0<a^2<F(\infty)$ such that $U(1)$ spontaneous symmetry breaking is ensured, the last two terms in the energy integral combine as a boundary contribution
that is proportional to the magnetic flux $\Phi_M$. Thus, finite-energy configurations comply with the inequality $E\geq \frac{1}{2} e a^2 |\Phi_M|$, and the Bogomolny bound is saturated if and only if the first-order self-duality equations
\beq
B=\pm\frac{e}{2}\left(a^2-F(|\varphi|^2)\right)\hspace{0.6cm} ,\hspace{0.6cm} D_1\varphi\pm iD_2\varphi=0\label{bog2}
\eeq
are satisfied.  For radially symmetric fields $\varphi(r,\theta)=f(r) e^{i n\theta}$, $r A_\theta=n\beta(r)$ in the topological sector with winding number $n$, the PDE system (\ref{bog2}) becomes the ODE system of coupled equations
\beq
\frac{n}{r}\frac{d\beta}{dr}=\pm \frac{e}{2}\left[ a^2-F(f^2)\right]\hspace{0.6cm} ,\hspace{0.6cm}
\frac{d f}{dr}=\pm \frac{n}{r} (1- e \beta) f . \label{bog55}
\eeq
Boundary conditions on the solutions are dictated by regularity at the origin and energy finiteness:
$f(0)=0$, $\beta(0)=0$, $f(\infty)=v$ and $\beta(\infty)=\frac{1}{e}$. $v^2$ is a zero of the potential, $F(v^2)=a^2$. The solutions are vortices and anti-vortices with magnetic flux $e \Phi_M=2 \pi n$ and energy per unit length $E=\pi|n| a^2$. Beyond the radially symmetric case, an index theorem calculation gives the dimension of the moduli space of solutions of (\ref{bog2}) in each topological sector, see \cite{lohe,wein}. The space of linear deformations of a vortex that preserve the self-duality equations is the kernel of the differential operator
\bdm
{\cal D}=\left(\begin{array}{cccc}
\partial_1+ e A_2&-\partial_2+ e A_1&e\varphi_2&e\varphi_1\\
\partial_2-e A_1&\partial_1+ e A_2&-e\varphi_1&e\varphi_2\\
e g(|\varphi|^2) \varphi_1&e g(|\varphi|^2) \varphi_2&-\partial_2&\partial_1\\
0&0&\partial_1&\partial_2
\end{array}\right)
\edm
while, by introducing the auxiliary operator
\bdm
{\cal P}=\left(\begin{array}{cccc}
e \varphi_1&e \varphi_2&-\partial_2&\partial_1\\
e \varphi_2&-e \varphi_1&\partial_1&\partial_2\\
0&0&e \varphi_2&e \varphi_1\\
0&0&-e \varphi_1&e \varphi_2
\end{array}\right)
\edm
it is not difficult to show that ${\rm ker}\,{\cal P\,D^\dagger}=\{0\}$, and hence ${\rm ker}\,{\cal D^\dagger}=\{0\}$. This vanishing theorem and the supersymmetric pairing of the non-zero eigenvalues of the Laplacians associated to ${\cal D}$ and ${\cal D}^\dagger$, give the dimension of the moduli space as
\bdm
{\rm dim}\,{\rm ker}{\cal D}={\rm ind}\,{\rm ker}{\cal D}={\rm Tr}\left\{\frac{M^2}{{\cal D}^\dagger{\cal D}+M^2}\right\}-{\rm Tr}\left\{\frac{M^2}{{\cal D}{\cal D}^\dagger+M^2}\right\}=\frac{e}{\pi} \Phi_M=2 n,
\edm
where the result comes from trace evaluation in the limit $M^2\rightarrow\infty$.  The interpretation, \cite{jatau,wang}, is that the Higgs field of the general solution with winding number $n$ has $|n|$ zeroes along the plane, and the $2|n|$ parameters of the solution correspond to the coordinates of these zeroes.  The self-dual defects are thus non-interacting solitons at the verge between two superconducting regimes. Rewriting the coupling in (\ref{sdpot}) as $\lambda e^2$, one has that for $\lambda <1 $ the mass of the
vector boson would be greater than that of the Higgs boson and long-range inter-vortex forces would be attractive, thus one deals with type I superconductivity, whereas in the opposite $\lambda\geq 1$ case, the vortices would repel each other belonging to a type II superconductivity phase.

\subsection{Self-dual vortices in the south chart}

In the gauged non-linear sigma model on the sphere, the potential energy density, together with the metric function, are chosen to be
\beq
g_S(|\phi|^2)=\frac{4\rho^4}{(\rho^2+|\phi|^2)^2},\hspace{1cm} U_S(|\phi|^2)=\frac{e^2}{8}\left(\frac{4 \rho^2 |\phi|^2}{\rho^2+|\phi|^2} -a^2\right)^2, \label{gUsur}
\eeq
in order to find the Lagrangian (\ref{lagphi}) at the critical self-dual point in the south chart. It is required that $0<a^2<4\rho^2$ whereas $U_S(|\phi|^2)$ vanishes along the vacuum orbit
\bdm
|\phi|^2=v^2=\frac{\rho^2 a^2}{4\rho^2-a^2},
\edm
see Figure 1(left). The Higgs mechanism occurs and provides identical masses to the vector and the Higgs bosons:
\[
m_A^2=m_H^2=\frac{ \rho^2 a^2 e^2}{\rho^2+v^2}=\frac{4\rho^2-a^2}{4\rho^2}e^2a^2=m^2 \quad .
\]
\begin{figure}[ht]
\centerline{\includegraphics[height=2.5cm]{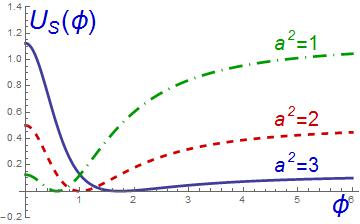}\hspace{2cm}
\includegraphics[height=2.5cm]{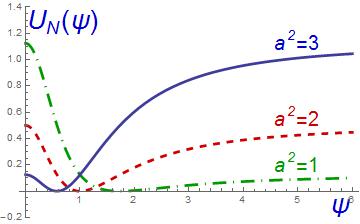}\hspace{0.3cm}}
\caption{\small Scalar potential on the $\phi$-chart (left) and $\psi$-chart (right) for $\rho=1$ and several values of the parameter $a^2$.}
\end{figure}

The Bogomolny equations are
\beq
B=\pm\frac{e}{2}\left(a^2-\frac{4 \rho^2 |\phi|^2}{\rho^2+|\phi|^2}\right),\hspace{2cm}D_1\phi\pm i D_2\phi=0\label{eqphi2}
\eeq
and the vortex energy per unit length is $E=\pi|n| a^2$ for a winding number $n$. The scalar field maps the center of the vortices to the south-pole of the sphere, whereas the image of the boundary circle of the plane $r\to +\infty$ is the parallel circle $|\phi|=v$ in the south-chart of the sphere. The radially symmetric configurations $\phi(r,\theta)=f(r)e^{in\theta}$ and $rA_\theta=n\beta(r)$ are BPS $\phi$-vortex solutions if the functions $f(r)$ and $\beta(r)$ solve the ODE system (\ref{bog55}) for the choice of $F(\vert\phi\vert^2)$ as a primitive of the function $g_S(|\phi|^2)$ written in formula (\ref{gUsur}). The solutions have been determined numerically by a standard shooting procedure for configurations with vorticity $n=1,2,3,4$ and are shown in  Figure 2, together with the energy density $\epsilon(r)$ and the magnetic field $B(r)$.

\begin{figure}[ht]
\centerline{\includegraphics[height=2.4cm]{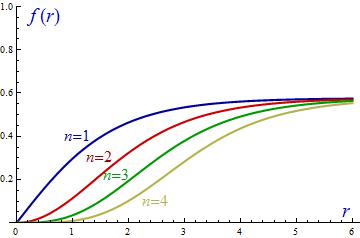}
\includegraphics[height=2.4cm]{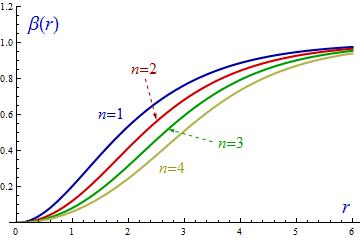}
\includegraphics[height=2.4cm]{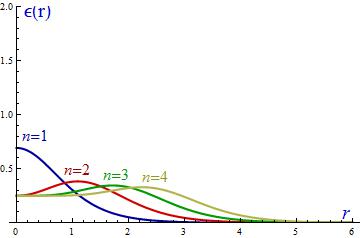}
\includegraphics[height=2.4cm]{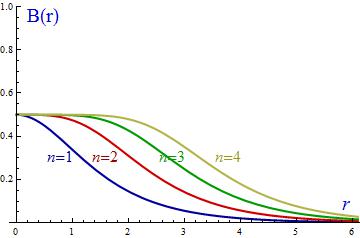}}
\caption{\small Profiles of radially symmetric $\phi$-vortices for several values of the vorticity $n$ and $\rho=a=1$: from left to right, the function $f(r)$, the function $\beta(r)$, the energy density $\epsilon(r)$ and the magnetic field $B(r)$.}
\end{figure}

Figure 3 is a graphical representation of the $n=1$ and $n=2$ self-dual vortices for $\rho=1=a$ using the original valued-on-the sphere field $\Phi=(\Phi_1,\Phi_2,\Phi_3)$, which is plotted as a vector of modulus $\rho=1$ on each point of the spatial plane. For pictorial reasons the projections $\Phi_1$ and $\Phi_2$ of the vector $\Phi$ are represented on the spatial axes in order to represent a five-dimensional space in a three-dimensional picture. For instance, at the center of the vortex, where the stereographic coordinates vanish, the isovector $\Phi$ points towards the south pole $\Phi_3=-1$. As the distance from the vortex center grows the arrow configuration follows the twists demanded by the solution (\ref{invstp}), always standing below the \lq\lq vacuum\rq\rq parallel. At infinite distances from the origin the isovector field reaches this circle winding once or twice around the vacuum parallel.

\begin{figure}[ht]
\centerline{\includegraphics[height=3.cm]{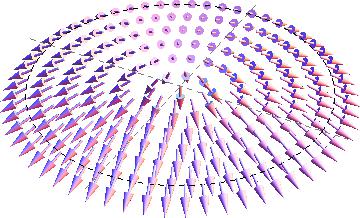} \hspace{1cm}
\includegraphics[height=3.cm]{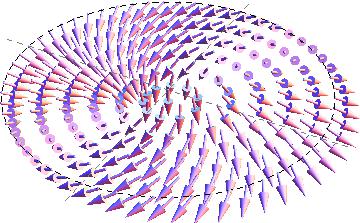}}
\caption{\small Graphical representation of the isospin vortex field $\Phi=(\Phi_1(x_1,x_2), \Phi_2(x_1,x_2), \Phi_3(x_1,x_2))$ for the $\phi$-vortices with $n=1$ (left) and $n=2$ (right) for $\rho=a=1$ in the spatial plane.}
\end{figure}

\subsection{Self-dual vortices in the north chart}

The field redefinition $\psi^*=\frac{\rho^2}{\phi} $ applied in formula (\ref{gUsur}), together with the subsequent changes (\ref{cambioderi}), leads to the self-dual Lagrangian (\ref{lagpsi}) in the $\psi$-chart. Thus, in this north chart we find:
\beq
g_N(|\psi|^2)=\frac{4\rho^4}{(\rho^2+|\psi|^2)^2}\hspace{1cm}U_N(|\psi|^2)=\frac{e^2}{8}\left(\frac{4 \rho^2 |\psi|^2}{\rho^2+|\psi|^2} -(4\rho^2-a^2)\right)^2, \label{potnc}
\eeq
see Figure 1(right). The north-chart Bogomolny equations are
\beq
B=\pm\frac{e}{2}\left(4\rho^2-a^2-\frac{4 \rho^2 |\psi|^2}{\rho^2+|\psi|^2}\right),\hspace{1cm}
D_1\psi\pm i D_2\psi=0\label{eqpsi2}
\eeq
and the BPS energy of vortices and anti-vortices per unit length is $E=\pi|n| (4\rho^2-a^2)$. The point of the scalar configuration space corresponding to the center of the defects is now the north-pole, whereas the fields at large distances reach the parallel
\bdm
|\psi|^2=w^2 = \frac{\rho^2(4\rho^2-a^2)}{a^2}
\edm
in the $\psi$-chart. Because $w^2=\frac{\rho^4}{v^2}$, this is the same parallel that we found in the $\phi$-chart and represents the global vacuum orbit of the theory. Moreover, the Higgs
mechanism in this chart produces the same masses for the vector and Higgs bosons as in the south chart:
\[
m^2_H=m_A^2=\frac{4\rho^2-a^2}{\rho^2+w^2}\rho^2 e^2=\frac{4\rho^2-a^2}{4\rho^2}e^2a^2=m^2 \quad .
\]
The system is thus globally well defined.

The vacuum orbit divides the sphere into two disjoint skullcaps, the scalar field of $\phi$- and $\psi$- vortices or antivortices respectively taking values in the south and north ones. Radially symmetric configurations in this second chart $\psi(r,\theta)=\widetilde{f}(r)e^{in\theta}$ and $rA_\theta=n\widetilde{\beta}(r)$ are BPS-vortex solutions living in the north skullcap if $\widetilde{f}(r)$ and $\widetilde{\beta}(r)$ solve the ODE system (\ref{bog55}) for the $F(\vert\psi\vert^2)$ function appearing in (\ref{potnc}). The system is again solved numerically and the $\widetilde{f}(r)$ and $\widetilde{\beta}(r)$ profiles, together with the energy density $\widetilde{\epsilon}(r)$ and the magnetic field $\widetilde{B}(r)$, are respectively displayed in Figure 4 for the $\psi$-vortices with vorticity $n=1,2,3,4$.

\begin{figure}[ht]
\centerline{\includegraphics[height=2.4cm]{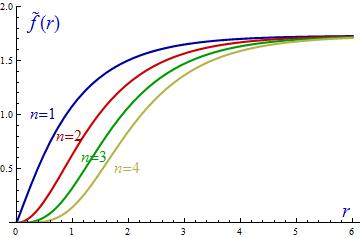}
\includegraphics[height=2.4cm]{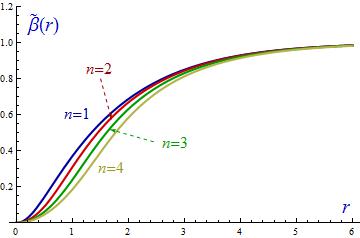}
\includegraphics[height=2.4cm]{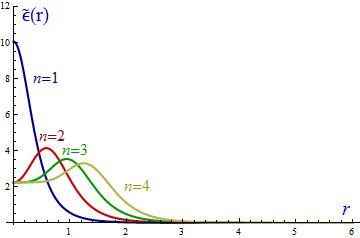}
\includegraphics[height=2.4cm]{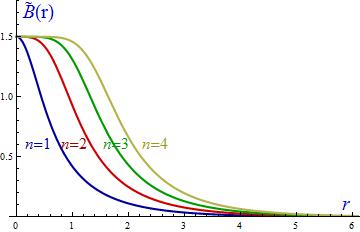}}
\caption{\small Profiles of radially symmetric $\psi$-vortices for several values of the vorticity $n$ and $a=1$: from left to right the function $\widetilde{f}(r)$, the function $\widetilde{\beta}(r)$, the energy density $\widetilde{\epsilon}(r)$ and the magnetic field $\widetilde{B}(r)$.}
\end{figure}

Figure 5 is a graphical representation of the $\Phi$-field for these solution with $n=1$ and $n=2$. In this case the isovector field point towards the north pole $\Phi_3=1$ at the origin $x_1=0,x_2=0)$ of the plane, follows the twists required by the numerical solution always standing in the north hemisphere
until reaching the $\Phi_3=\frac{1}{2}$ parallel, from above in $\mathbb{S}^2$, asymptotically in $\mathbb{R}^2$. The vacuum parallel is then wrapped once or twice by the isovector field at spatial infinity.

\begin{figure}[ht]
\centerline{\includegraphics[height=3.cm]{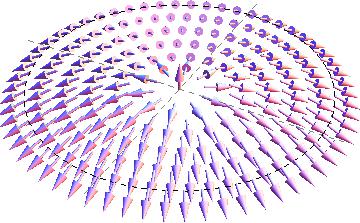} \hspace{1cm}
\includegraphics[height=3.cm]{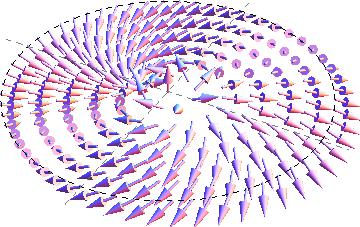}}
\caption{\small Graphical representation of the isospin vortex field $\Phi=(\Phi_1(x_1,x_2),\Phi_2(x_1,x_2),\Phi_3(x_1,x_2))$ for the $\psi$-vortices with $n=1$ (left) and $n=2$ (right) for $a=1$ in the spatial plane.}
\end{figure}

In sum, there exist two species of BPS vortices whose energies are respectively $E=\pi |n| a^2$ and $E=\pi |n| (4\rho^2-a^2)$ for configurations with $n$-vorticity. From Figures 2 and 4 we observe that for $a=1$ the first type of vortices supports thick profiles, whereas vortices of the second species are thin; the energy per unit length of thick/thin vortices is less/more concentrated in the plane. The difference in vortex energy density width, is more pronounced when the vacuum parallel $|\phi|^2=v^2$ approaches the North Pole, and disappears when the vacuum orbit is the Equator, $a^2=2\rho^2$, and thick vortices become thin and viceversa if the vacuum orbit lies in the south hemisphere, $a^2>2\rho^2$.

A subtle point is the following: although $\phi$- and $\psi$- vortices with the same winding number live in the same topological sector, both of them are separately stable. The higher-energy per unit length vortices do not decay to the lower-energy per unit length ones. The scalar field at their centres is respectively the North or the South Pole, depending on the species. In fact, BPS vortices of different species saturate either the $\phi$- or $\psi$- Bogomolny bounds, which are valid on either the south or the north charts. The circumstance that each type of vortex carries a different energy per unit length seems a bit puzzling: the BPS bounds coming from the $\phi$- and $\psi$- field Bogomolny splittings are, respectively, $E\geq \frac{e}{2} a^2 \Phi_M$ and $E\geq \frac{e}{2} (4\rho^2-a^2) \Phi_M$. Thus, if $2\rho^2 \neq a^2$, the energy of $\psi$-vortices contradicts the $\phi$-bound, and viceversa. In fact, however, the contradiction is only apparent and  is due, contrary to what happens in the usual Abelian Higgs model, to the fact that finite-energy configurations can have isolated points where the scalar field goes to infinity. Recall that the target space is $\mathbb{C}\mathbb{P}^1=\mathbb{C} \cup \{\infty\}$. A radially symmetric configuration $\phi(r,\theta)=f(r) e^{i n_\infty \theta}$ such that $\lim_{r\rightarrow 0}f(r)=\infty$ is therefore admissible. In this case, the antisymmetric sum of double derivatives of $\ln\phi$ produces a new contribution to the $R$ term of the Bogomolny splitting of the form
\bdm
\Delta R=\frac{i}{2} F(|\phi|^2)\varepsilon_{ij} \partial_i\partial_j\ln\phi=-\pi n_\infty F(|\phi|^2) \delta^{(2)}(\vec{x}).
\edm
The $\phi$-field Bogomolny bound in the complex plane plus the infinity point becomes
\bdm
{\displaystyle E\geq \pi \left|n_0 a^2-n_\infty(4\rho^2-a^2)\right|}
\edm
because $\lim_{\vert\phi\vert\to +\infty}\left(\frac{4 \rho^2 |\phi|^2}{\rho^2+|\phi|^2}-a^2\right)=4\rho^2-a^2$. In the previous formula, $n_0$ denotes the number of zeroes of $\phi$ in $\mathbb{C}$, and therefore the total winding number is $n=n_0+n_\infty$. This is the global BPS bound that encompasses the bounds in both charts. If $n_\infty=0$, there are vortices only in the south chart, $n=n_0$, and we find the BPS $\phi$-bound. $n_0=0$ means that $n_\infty=n$, which is tantamount to the existence
of $\psi$-vortices with vorticity $n$ in the south chart, in perfect agreement with the BPS $\psi$-bound. The $\phi$- and $\psi$- Bogomolny bounds are the two local forms of the global Bogomolny bound.

We conclude that the gauged massive non-linear $S^2$-sigma model admits stable self-dual solutions in the form of either $\phi$- or $\psi$- vortices. One might wonder if there are also solutions given by the symbiosis of defects of both species. When the change of variables $\phi=\frac{\rho^2}{\psi^*} $ is applied to a configuration satisfying (\ref{eqphi2}) with the plus sign, we obtain a configuration that satisfies (\ref{eqpsi2}) with the minus sign, and viceversa. One might think that a superposition of a $\phi$-vortex with a $\psi$-antivortex could be a self-dual configuration of the theory, but in this case the total winding number vanishes, there is no topological obstruction to ensure stability,
and these configurations are not self-dual. Notice, however, that for $n_0=1,n_\infty=-1$  the global Bogomolny bound gives an energy per unit length $E=4\pi\rho^2$, which is the area of the sphere and coincides with the energy of a fundamental $\mathbb{CP}^1$ lump in the class $N=1$ of the homotopy group $\pi_2(\mathbb{S}^2)$. In fact, as it is understood in \cite{nivi}, a different topological interpretation of these $\phi$-vortex $\psi$-antivortex pairs is possible; in it,  they appear as the basic constituents of $\mathbb{CP}^1$ lumps.  On the other hand, the superposition of a $\phi$-vortex with a $\psi$-vortex could also be a non self-dual solution of the Euler-Lagrange equations, but this can only be decided by a numerical evaluation of the interaction energy  \`{a} la Jacobs-Rebbi \cite{jare}, which we have so far not undertaken. In any case, given that the Higgs mechanism gives mass $m$ to the elementary particles, the interaction energy among defects located at distances of order  $R\gg \frac{1}{m}$ is small, $E_{\rm int}\simeq e^{- 2 m R}$ \cite{vish}, and mixed configurations of widely separated $\phi$- and $\psi$- defects can always be considered good approximate solutions to the Euler-Lagrange equations.

\subsection{More globally self-dual models}

We have built the non-linear Abelian-Higgs sigma model using the natural round metric on the sphere, but once the gauge field has been introduced and the potential energy density has been chosen, the symmetry is reduced from the $\mathbb{O}(3)$ group to the $\mathbb{U}(1)$ subgroup of rotations around the $\Phi_3$-axis. There is nothing to prevent us from considering models on the sphere with other metrics as long as they respect this reduced symmetry. We must mention that in a more sophisticated context, ${\cal N}=2$ supersymmetric gauge theories where the vacuum moduli is $\mathbb{C}\mathbb{P}^1$ with some singularity,  other K\"{a}hler metrics than the standard Fubini-Study metric have been considered, see Reference \cite{nivi1} to find  interesting BPS defects characterized as fractional vortices. There is, however, an appealing feature of the round metric that we would like to preserve: this is the fact that it gives rise to globally self-dual models, i.e. compatible self-dual structures arise in each chart of the sphere. By this we mean that the self-duality equations on both charts have almost the same form, the only difference being the value of $W(0)$. Next we shall briefly describe a possible way to produce a number of other non-linear sigma models with this type of global self-duality. Working on the south chart, the idea is to take a dimensionless function $f(|\phi|^2)$ such that $f(0)=0$, $f^\prime(|\phi|^2)\geq 0, \forall\vert\phi\vert$, and $f(\infty)=q^2\leq\infty$. We then define the $F$ part of the potential as $F(|\phi|^2)=\rho^2\left(f(|\phi|^2)-f(\frac{\rho^4}{|\phi|^2})+q^2\right)$ in such a way that $F(0)=0$, as it should be, and $F(\infty)=2\rho^2 q^2$ is finite. Therefore, the metric and potential on the south chart are
\bdm
g_S(|\phi|^2)=\rho^2\left[f^\prime(|\phi|^2)+\frac{\rho^4}{|\phi|^4} f^\prime\Big(\frac{\rho^4}{|\phi|^2}\Big)\right],\hspace{1cm}U_S(|\phi|^2)=\frac{e^2}{8}\left[F(|\phi|^2)-a^2\right]^2,
\edm
whereas the change of fields (\ref{cambioderi}) gives the following metric and potential for the $\psi$ field:
\bdm
g_N(|\psi|^2)=\rho^2\left[f^\prime(|\psi|^2)+\frac{\rho^4}{|\psi|^4} f^\prime\Big(\frac{\rho^4}{|\psi|^2}\Big)\right]
,\hspace{0.2cm}U_N(|\psi|^2)=\frac{e^2}{8}\left[F(|\psi|^2)-(2\rho^2 q^2-a^2)\right]^2.
\edm
Self-duality in the two charts is thus apparent. The energies per unit length of $\phi$-
and $\psi$- vortices or antivortices with winding number $n$ are
respectively $E=\pi |n| a^2$ and $E=\pi |n|(2\rho^2 q^2- a^2)$, and the
phenomenology is analogous to what has been described for the model with the
round metric. The case of the function $f(|\phi|^2)=\ln\frac{\rho^2+2 |\phi|^2}{\rho^2+|\phi|^2}$ gives an interesting example where double self-duality combines with a logarithmic potential.

\section{Dielectric functions and BPS topological defects in the gauged $\mathbb{S}^2$-sigma model}

\subsection{The Lagrangian and the Bogomolny splitting}
The kinetic energy density for the gauge field in the Lagrangians (\ref{lagphi}) and (\ref{lagpsi}) that we have been dealing with in the previous section is given by the standard Maxwell term. This is the canonical choice when the model is intended to describe the interactions of fundamental quanta in vacuo, but there are many physical systems in which the Abelian-Higgs model plays the r$\hat{\rm o}$le of an effective theory ruling the dynamics of the excitations of some background medium. For this sort of application, it may be the case that the minimal Maxwell term has to be supplemented with a dielectric function that will account for the enhancement or screening of the forces among quanta due to the polarization of the underlying condensate. The $\mathbb{U}(1)$ gauge symmetry requires that the dielectric factor should depend only on the scalar field modulus, and hence the Lagrangian of Section 3 changes to the form
\begin{equation}
{\cal L}=-\frac{1}{4} H(|\varphi|^2) F_{\mu\nu} F^{\mu\nu} +\frac{1}{2} g(|\varphi|^2) D_\mu \varphi^* D^\mu \varphi-\frac{1}{2} W^2(|\varphi|^2) \quad . \label{dielagr}
\end{equation}
Models of this type where $H(|\varphi|^2)$ is a positive definite function admit first-order Bogomolny equations, see e.g. \cite{leenam, bhsm, bchm,Fuerguil}. The  arrangement
\begin{eqnarray}
E/L&=&\int d^2 x \left\{ \frac{1}{2} \left(\sqrt{H(|\varphi|^2)}B\pm W(|\varphi|^2)\right)^2+\frac{1}{2}g(|\varphi|^2) |D_1\varphi\mp i D_2\varphi|^2\right. \nonumber\\ &\mp& \left.(\sqrt{H(|\varphi|^2)}W(|\varphi|^2) B+ R\right\} \label{dielbogs}
\end{eqnarray}
leads, together with the choice of the potential as
\beq
U(\varphi)=\frac{1}{2} W^2(|\varphi|^2)=\frac{e^2}{8}\frac{(F(|\varphi|^2)-a^2)^2}{H(|\varphi|^2)} \quad ,
\eeq
to self-duality equations of the form
\beq
B=\pm\frac{e}{2}\frac{\left(a^2-F(|\varphi|^2)\right)}{H(|\varphi|^2)},\hspace{0.6cm} ,\hspace{0.6cm} D_1\varphi\pm iD_2\varphi=0
\eeq
whose solutions are vortices and antivortices of energy density $E=\pi |n| a^2$ for winding number $n$.

\subsection{BPS vortices in the gauged $\mathbb{S}^2$-sigma model with dielectric function and circular vacuum orbit}

In practice, the dielectric function can be chosen in many different ways that can generate a broad variety of vortex profiles. For the non-linear sigma model on the sphere, a natural option is to use a function $H(|\varphi|^2)$, leading to regular vortices with a similar structure in both the south and north charts, as indeed happened in the minimal model without dielectric function. To achieve this behaviour we choose the south chart dielectric function and self-dual potential in the form
\bdm
H_S(|\phi|^2)=\frac{c_0 |\phi|^2+c_1 \rho^2}{b_0 |\phi|^2+b_1 \rho^2},\hspace{1cm}  U_S(|\phi|^2)=\frac{e^2}{8}\frac{b_0 |\phi|^2+b_1\rho^2}{c_0 |\phi|^2+c_1\rho^2}\left(\frac{4 \rho^2 |\phi|^2}{\rho^2+|\phi|^2} -a^2\right)^2,
\edm
where $b_0, b_1,c_0,c_1$ are strictly positive real numbers.

\begin{figure}[ht]
\centerline{\includegraphics[height=3.cm]{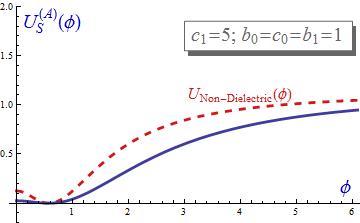} \hspace{1cm} \includegraphics[height=3.cm]{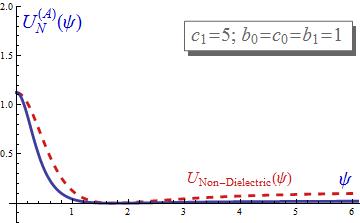}}
\centerline{\includegraphics[height=3.cm]{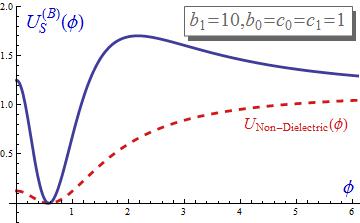} \hspace{1cm} \includegraphics[height=3.cm]{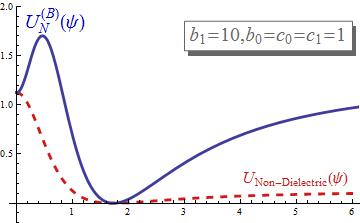}}
\caption{\small Scalar potentials in the south and north charts for cases A ($c_1=5$, $b_0=c_0=b_1=1$) and B ($b_1=10$, $b_0=c_0=c_1=1$) for the values $a=\rho=e=1$.}
\end{figure}

The vacuum orbit occurs in these models along the circle $|\phi|=\frac{a \rho}{\sqrt{4\rho^2-a^2}}$, and there is always a critical point of the potential at $\phi=0$ that can be tuned to be a local minimum or maximum with a suitable selection of the parameters. In particular, if $c_0=b_0$ and $c_1=b_1$, the function $H$ becomes  unity and we recover the system where the Nitta-Vinci vortices arise. Changing coordinates to the north chart, the metric mutates to the known round $g_N(|\psi|^2)$ metric on $\mathbb{S}^2$, whereas the dielectric function and potential energy density become
\bdm
H_N(|\psi|^2)=\frac{c_1 |\psi|^2+c_0\rho^2}{b_1 |\psi|^2+b_0\rho^2},\hspace{0.3cm}  U_N(|\psi|^2)=\frac{e^2}{8}\frac{b_1 |\psi|^2+b_0\rho^2}{c_1 |\psi|^2+c_0\rho^2}\left(\frac{4 \rho^2 |\psi|^2}{\rho^2+|\psi|^2} -(4\rho^2-a^2)\right)^2.
\edm
The duality between the theories with $(b_0,b_1,c_0,c_1; a^2)$ and $(b_1,b_0,c_1,c_0; 4\rho^2-a^2)$ is thus patently clear. The regular vortices of the south (north) chart in one theory are akin to the regular vortices appearing in the north (south) chart in the other. In particular, the energy per unit length of the vortices in the north chart is: $E=\pi(4\rho^2-a^2)\vert n\vert$. The same argument developed in the microscopic scenario of the previous section regarding the existence of a global Bogomolny bound works here.

The ODEs giving the radially symmetric defects are of the generic form
\beq
\frac{n}{r}\frac{d\beta}{dr}=\pm \frac{e}{2}\frac{\left[ a^2-F(f^2)\right]}{H(f^2)}\hspace{0.6cm} ,\hspace{0.6cm}
\frac{d f}{dr}=\pm \frac{n}{r} (1- e \beta) f
\eeq
where, of course, south- or north- variables and $F$ and $H$ functions have to be appropriately substituted and finite-energy boundary conditions respected. The energy density per unit length and the magnetic field for these solutions are:
\begin{eqnarray*}
{\cal E}(r) &=& \frac{n^2}{r^2}g[f^2(r)]f^2(r)[1-e\beta(r)]^2+\frac{e^2}{4}\frac{(a^2-F[f^2(r) ])^2}{H[f^2(r)]}   \\ B(r)&=&\frac{e^2}{2}\frac{a^2-F[f^2(r)]}{H[f^2(r)])}\quad .
\end{eqnarray*}
As an illustration, here we shall present the solutions for two cases, let us call them A and B, where the parameters are chosen to be $\rho=a=b_0=b_1=c_0=1, c_1=5$ in Case A, and $\rho=a=b_0=c_0=c_1=1, b_1=10$ in Case B. The potential energy densities for these parameters are plotted in Figure 6 (blue lines), where a comparison with the potentials (red lines) giving self-duality in the microscopic, non dielectric, case is also offered.  Observation of these graphics reveals to us that the potentials in case A are closer to the non-dielectric self-dual potentials than the potentials in case B  in both charts. In the second model, one can observe that there is a local maximum of $U$ at $\phi=0$ in the south chart, but the potential in the south
chart shows a local minimum at $\psi=0$. In the next section we shall examine a new case where the parameter $b_1$ vanishes (case C).

\begin{figure}[ht]
\centerline{\includegraphics[height=2.4cm]{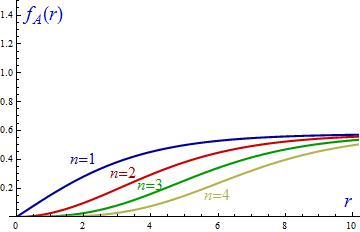}\hspace{0.1cm}
\includegraphics[height=2.4cm]{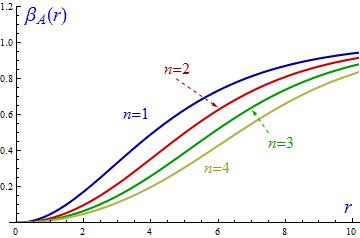}\hspace{0.1cm}
\includegraphics[height=2.4cm]{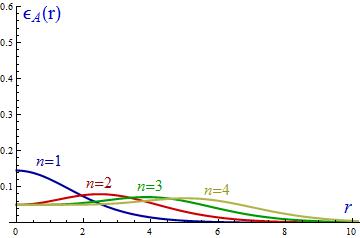}\hspace{0.1cm}
\includegraphics[height=2.4cm]{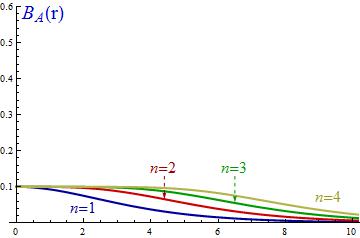}}
\caption{{\small Profiles of the radially symmetric $\phi$-vortices for several values of the vorticity $n$ in case A: from left to right the function $f(r)$, the function $\beta(r)$, the energy density $\epsilon(r)$ and the magnetic field $B(r)$.}}
\end{figure}

\begin{figure}[ht]
\centerline{\includegraphics[height=2.4cm]{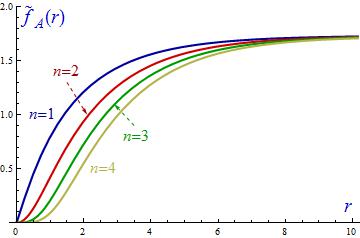}\hspace{0.1cm}
\includegraphics[height=2.4cm]{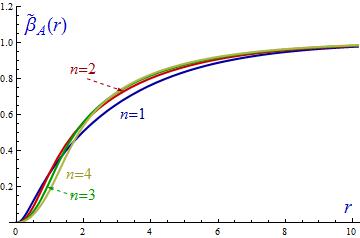}\hspace{0.1cm}
\includegraphics[height=2.4cm]{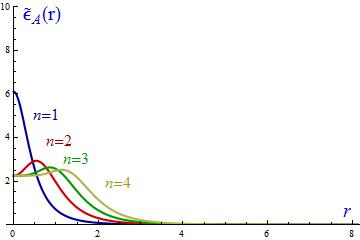}\hspace{0.1cm}
\includegraphics[height=2.4cm]{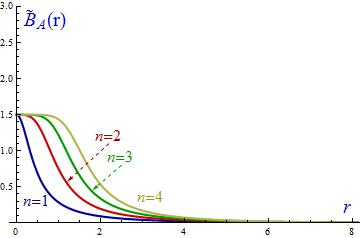}}
\caption{{\small Profiles of the radially symmetric $\psi$-vortices for several values of the vorticity $n$ in case A: from left to right the function $\widetilde{f}(r)$, the function $\widetilde{\beta}(r)$, the energy density $\widetilde{\epsilon}(r)$ and the magnetic field $\widetilde{B}(r)$.}}
\end{figure}

Figures 7 to 10 show the specific scalar and vector boson field profiles, as well as the energy densities and the magnetic fields, of the radially symmetric solutions for cases A and B with winding numbers increasing from 1 to 4. Again, these magnitudes in  case A are close to the field profiles and density energies of the Nitta-Vinci vortices. Interesting new features arise in case B profiles: namely, the magnetic field presents a local minimum at $\psi=0$ in the north chart even for solutions with vorticity $n=1$. Thus, the maximum values of the magnetic field are attained at a ring in the plane enclosing the origin, a configuration that resembles the self-dual planar Chern-Simons-Higgs vortices. Indeed this parallelism is more direct for the case in which the parameter $b_1$ vanishes as we shall show in the next subsection.

\begin{figure}[ht]
\centerline{\includegraphics[height=2.4cm]{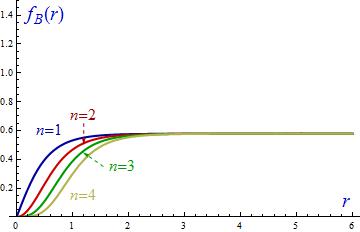}\hspace{0.1cm}
\includegraphics[height=2.4cm]{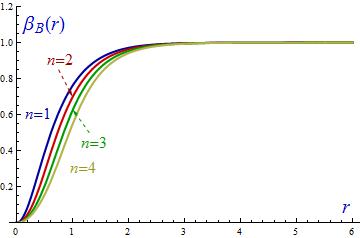}\hspace{0.1cm}
\includegraphics[height=2.4cm]{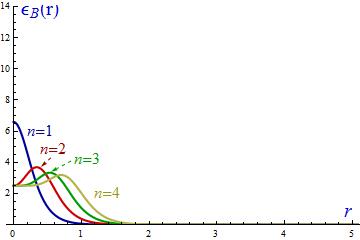}\hspace{0.1cm}
\includegraphics[height=2.4cm]{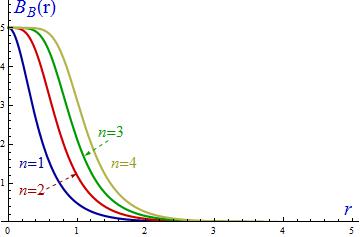}}
\caption{\small Profiles of the radially symmetric $\phi$-vortices for several values of the vorticity $n$ in case B: from left to right the function $f(r)$, the function $\beta(r)$, the energy density $\epsilon(r)$ and the magnetic field $B(r)$.}
\end{figure}
\begin{figure}[ht]
\centerline{\includegraphics[height=2.4cm]{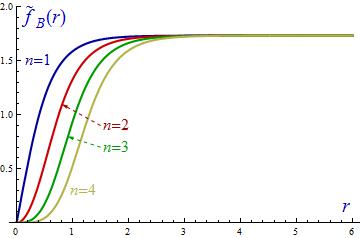}\hspace{0.1cm}
\includegraphics[height=2.4cm]{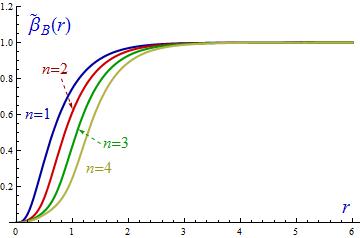}\hspace{0.1cm}
\includegraphics[height=2.4cm]{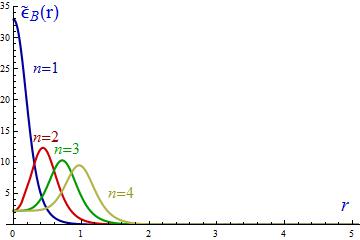}\hspace{0.1cm}
\includegraphics[height=2.4cm]{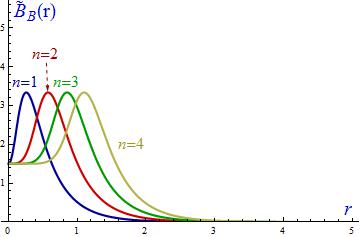}}
\caption{\small Profiles of the radially symmetric $\psi$-vortices for several values of the vorticity $n$ in case B: from left to right the function $\widetilde{f}(r)$, the function $\widetilde{\beta}(r)$, the energy density $\widetilde{\epsilon}(r)$ and the magnetic field $\widetilde{B}(r)$.}
\end{figure}

\subsection{Dielectric functions giving rise to non-connected moduli space of vacua}

New features appear in these self-dual models and a new structure of BPS solutions arise when the dielectric function is selected with $b_1=0$.

\subsubsection{Self-dual vortices in the south chart}

In this case we have in the south chart the following functions setting the dynamics:
\[
H_S(|\phi|^2)=\frac{c_0 |\phi|^2+c_1 \rho^2}{b_0 |\phi|^2},\hspace{1cm}  U_S(|\phi|^2)=\frac{e^2}{8}\frac{b_0 |\phi|^2}{c_0 |\phi|^2+c_1\rho^2}\left(\frac{4 \rho^2 |\phi|^2}{\rho^2+|\phi|^2} -a^2\right)^2.
\]

\begin{figure}[ht]
\centerline{\includegraphics[height=3cm]{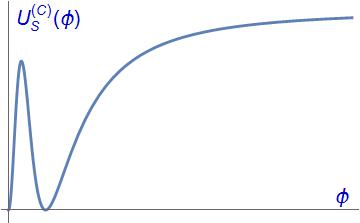} \hspace{1cm} \includegraphics[height=3cm]{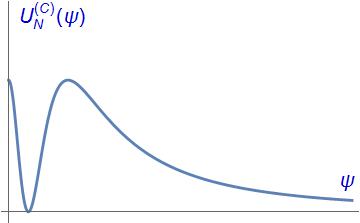}}
\caption{\small Scalar potentials in the south and north charts  $U_S^C(|\phi|)$ and $U_N^C(|\psi|)$ for the case C.}
\end{figure}

The behaviour of the potential $U_S(|\phi|^2)$ is displayed in Figure 11(left). The vacuum orbit is formed by the disjoint union of the south pole $\phi_0=0$ and the parallel $\vert\phi_a\vert=\frac{\rho a}{\sqrt{4\rho^2-a^2}}$,
\[
{\cal V}_S^{Df}=\Big\{ v_0=0 \Big\} \, \, \bigsqcup \, \, \Big\{ \vert v_a\vert=\frac{a\rho}{\sqrt{4\rho^2-a^2}}\Big\} \quad  ,
\]
while the asymptotic value of $U_S(|\phi|^2)$ remains finite. The action of the $\mathbb{U}(1)$ gauge group on the vacuum orbit provides a two point moduli space of vacua: ${\cal M}^{Df}= {\cal V}_S^{Df}/\mathbb{U}(1)=\{0,\frac{a \rho
}{\sqrt{4\rho^2-a^2}}\}$. Quantizing on the vacuum parallel produces a Higgs phase with massive vector and Higgs bosons. The novelty happens upon quantization on the south pole. There are no photon degrees of freedom but charged boson fluctuations of mass $m_\phi^2=\frac{e^2}{8}\cdot\frac{b_0 a^4}{c_1 \rho^2}$ emerge as fundamental quanta.

In order to search for BPS defects it is convenient to perform the redefinitions: $\phi=\rho\xi$, $v=\rho b=\rho\frac{a}{\sqrt{4\rho^2-a^2}}$. The potential energy density reads now:
\begin{equation}
U_S(\vert\xi\vert^2)=\frac{\lambda^2}{8}\cdot\frac{a^4}{b^4}\cdot\frac{\vert\xi\vert^2(\vert\xi\vert^2-b^2)^2}{(\vert\xi\vert^2+d^2)(\vert\xi\vert^2+1)^2} \quad , \quad \lambda^2=\frac{b_0}{c_0}e^2 \, \, , \, \, d^2=\frac{c_1}{c_0} \label{zator} \quad .
\end{equation}
The Bogomolny equations are thus:
\begin{equation}
B(x_1,x_2)=\pm\frac{\lambda^2}{2e}\cdot\frac{a^2}{b^2}\cdot\frac{\vert\xi\vert^2}{\vert\xi\vert^2+d^2}\cdot\frac{b^2-\vert\xi\vert^2}{\vert\xi\vert^2+1} \, \, , \, \, D_1\xi(x_1)\pm iD_2\xi(x_1,x_2)=0 \, \, \, . \label{bogeqcshdf2}
\end{equation}
With the appropriate asymptotic conditions the ODE system (\ref{bogeqcshdf2}) admits as solutions vortices, which are displayed in the Figure 12 for the parameter values $b_1=0$, $b_0=c_0=c_1=1$ and $\rho=a=e=1$. Notice that the magnetic field configuration consists of a ring around the vortex center where its value vanishes.

\begin{figure}[ht]
\centerline{\includegraphics[height=2.4cm]{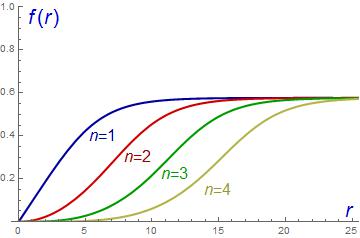}\hspace{0.1cm}
\includegraphics[height=2.4cm]{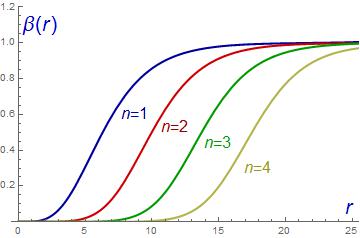}\hspace{0.1cm}
\includegraphics[height=2.4cm]{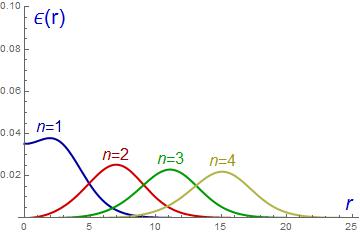}\hspace{0.1cm}
\includegraphics[height=2.4cm]{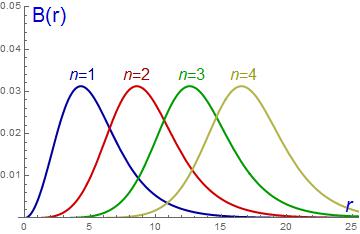}}
\caption{{\small Profiles of the radially symmetric $\phi$-vortices for several values of the vorticity $n$ in case C ($b_1=0$, $b_0=c_0=c_1=1$) for the parameters $\rho=a=e=1$: from left to right the function $f(r)$, the function $\beta(r)$, the energy density $\epsilon(r)$ and the magnetic field $B(r)$.}}
\end{figure}

\subsubsection{Self-dual vortices in the north chart}

In the north chart the following functions set the dynamics:
\[
H_N(|\psi|^2)=\frac{c_1 |\psi|^2+c_0 \rho^2}{b_0 \rho^2},\hspace{0.5cm}  U_N(|\psi|^2)=\frac{e^2}{8}\frac{b_0 \rho^2}{c_1 |\psi|^2+c_0\rho^2}\left(\frac{4 \rho^2 |\psi|^2}{\rho^2+|\psi|^2} -4\rho^2+a^2\right)^2.
\]
The behaviour of the potential $U_N(|\psi|^2)$ is shown in Figure 11(right), where we can notice that this function adopt a finite value at $|\psi|=0$ and vanishes asymptotically. Like in the south chart the vacuum orbit is formed by the disjoint union of the south pole $\psi_\infty=+\infty$, the parallel $\vert\psi_a\vert=\rho\frac{\sqrt{4\rho^2-a^2}}{a}=\frac{\rho}{b}$:
\[
{\cal V}_N^{Df}= \Big\{ w_\infty=+\infty  \Big\}\, \, \bigsqcup \, \, \Big\{ \vert w_a\vert=\rho\frac{\sqrt{4\rho^2-a^2}}{a} \Big\} \quad  .
\]
The action of the $\mathbb{U}(1)$ gauge group on the vacuum orbit provides a two point moduli space of vacua: ${\cal M}^{Df}= {\cal V}_N^{Df}/\mathbb{U}(1)=\{+\infty,\frac{\rho\sqrt{4\rho^2-a^2}}{a}\}$. Therefore, the vacuum manifold is globally defined and looks the same from
both charts: the south pole and the vacuum parallel. One may check that the particle spectrum in the symmetric and asymmetric phases are also identical in both charts. The field redefinition $\psi=\rho\chi$ shows that $\chi_\infty=+\infty$ and $\vert\chi_b\vert=1/b$ are the zeroes of the potential:
\[
U_N(\vert\chi\vert^2)=\frac{\lambda^2}{8}\cdot\frac{a^4}{b^4}\cdot\frac{1}{d^2\vert\chi\vert^2+1}\cdot \Big(\frac{1-b^2\vert\chi\vert^2}{1+\vert\chi\vert^2}\Big)^2 \, \, ,
\]
whereas the first-order ODE system becomes:
\begin{equation}
B(x_1,x_2)=\pm\frac{\lambda^2}{2e}\cdot\frac{a^2}{b^2}\cdot\frac{1}{d^2\vert\chi\vert^2+1}\cdot \frac{1-b^2\vert\chi\vert^2}{\vert\chi\vert^2+1} \, \, , \, \, D_1\chi(x_1)\pm iD_2\chi(x_1,x_2)=0 \, \, \, . \label{bogeqcshdf3}
\end{equation}
The behaviour of the radially symmetric solutions in this case follows the same pattern than the solutions depicted in the Figure 9.

\subsubsection{BPS magnetic domain walls}

The  structure of the moduli space ${\cal M}^{Df}$ allows for solutions of the self-duality equations (\ref{bogeqcshdf2}) other than vortices. Now we consider the axial gauge $A_1=0$ and looking forward for fields depending on $x_1$, such that $A_2=A_2(x_1)$ and $\xi=f(x_1)$ where we can take $f^*=f$ without loss of generality because of the global phase transformation. The energy per unit surface is given by
\[
E/A=\int \!\! dx_1 \left[ \frac{1}{2} H(\rho^2 f^2) \Big( \frac{dA_2}{dx_1} \Big)^2 + \frac{\rho^2}{2} g(\rho^2 f^2) \Big(\frac{\partial f}{\partial x_1}\Big)^2  + \frac{e^2\rho^2}{2} g(\rho^2 f^2) A_2^2 f^2 + U_N(\rho^2 f^2) \right]
\]
and the PDE system (\ref{bogeqcshdf2})
becomes a system of ordinary differential equations:
\begin{equation}
\frac{dA_2}{dx_1}(x_1)=\pm \frac{\lambda^2}{2 e}\cdot\frac{a^2}{b^2}\cdot\frac{f^2(x_1)(b^2-f^2(x_1))}{(f^2(x_1)+d^2)(f^2(x_1)+1)} \, , \, \, \, \frac{df}{dx_1}(x_1)\pm e A_2(x_1)f(x_1)=0 \label{sddw}\, \, .
\end{equation}
The ansatz $f(x_1)=e^{u(x_1)/2}$ together with the second equation in (\ref{sddw}) implies that $A_2(x_1)=\mp\frac{1}{2e}\frac{du}{dx_1}$ and
\[
E/A=\int \!\! dx_1 \left[ \frac{1}{2} H(\rho^2 e^u) \Big( \frac{dA_2}{dx_1} \Big)^2 + \frac{1}{8} g(\rho^2 e^u) \rho^2 e^u \Big(\frac{\partial u}{\partial x_1}\Big)^2 + \frac{e^2}{2} g(\rho^2 e^u) A_2^2 \rho^2 e^u + U_N(\rho^2 e^u) \right]
\]
One ends with the following second order ODE for $u$:
\begin{equation}
\frac{d^2u}{dx_1^2}=-2c\frac{e^u(b^2-e^u)}{(e^u+d^2)(e^u+1)} \quad , \quad c=\frac{\lambda^2 a^2}{b^2} \, \, . \label{mecqdw}
\end{equation}
This is no more than the motion equation for a particle of unit mass and position $u$ evolving in $x_1$-time under a gravitational field:
\[
V(u)= \left\{ \begin{array}{ll} \frac{2 c}{1-d^2}\Big[(b^2+d^2)\log\left(1+\frac{e^u}{d^2}\right)-(1+b^2)\log\left(1+e^u\right)\Big] & \mbox{ if } d^2\neq 1\\
2 c\Big[(b^2+1)\frac{e^u}{e^u+1}-\log\left(1+e^u\right)\Big] & \mbox{ if } d^2= 1 \end{array} \right. \quad .
\]
The integration constant is chosen such that: $V(-\infty)=0$. In any case $V(u)$ offers a first integral of equation (\ref{mecqdw}):
\[
\frac{1}{2}\left(\frac{du}{dx_1}\right)^2 + V(u)=I \qquad .
\]
The maximum value of the potential energy $V(u)$ is reached at $u=\log b^2$. Thus, for the choice $I=V(\log b^2)$, trajectories starting at $u(-\infty)=-\infty$ with velocity $\frac{du}{dx_1}(-\infty)=\sqrt{2V(\log b^2)}=-2 e A_2(-\infty)$ end at
$u(+\infty)=\log b^2$ with velocity zero, see Figure 13. These separatrices between particle motions bouncing back to $u(+\infty)=-\infty$ or running until $u(+\infty)=+\infty$ are given by the quadrature:
\begin{equation}
x_1-x_1^0=\int \, \frac{du}{\sqrt{2(V(\log b^2)-V(u))}} \label{magdw} \quad .
\end{equation}
Finally from the behaviour of the function $u(x)$ we can recover the form of the scalar and gauge fields $\xi(x_1)$ and $A_2(x_1)$, which are displayed in Figure 13, together with its energy density $\epsilon(x_1)$ and magnetic field $B_1(x_1)$ along the direction $x_1$.

\begin{figure}[ht]
\centerline{\includegraphics[height=2.2cm]{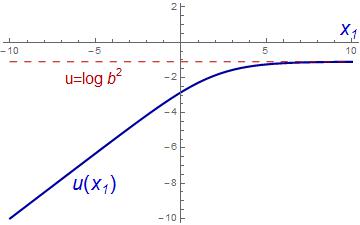}\hspace{0.1cm}
\includegraphics[height=2.2cm]{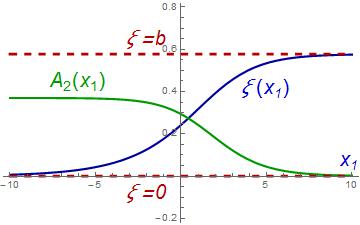}\hspace{0.1cm}
\includegraphics[height=2.2cm]{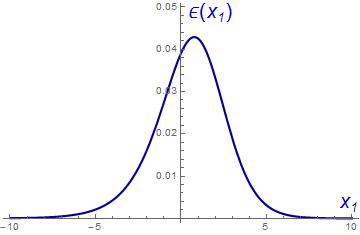}\hspace{0.1cm}
\includegraphics[height=2.2cm]{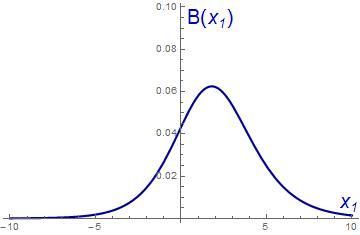}}
\caption{{\small Profiles of the BPS magnetic domain walls for the case C ($b_1=0$ and $b_0=c_0=c_1=1$ with $e=\rho=a=1$): from left to right the function $u(x_1)$, the functions $\xi(x_1)$ and $A_2(x_1)$, the energy density $\epsilon(x_1)$ and the magnetic field $B(x_1)$ per area unit.}}
\end{figure}

Assuming the system defined in an square of very large but finite area $A$ on the $x_2:x_3$-plane these kink-like BPS solitons interpolating between $v_0=0$ and a point in the circle $\vert v_a\vert=\frac{\rho a}{\sqrt{4\rho^2-a^2}}$ along the $x_1$ axis have finite tension, energy per unit surface:
\[
E/A=\frac{1}{2}e a^2 \left[A_2(+\infty)-A_2(-\infty)\right]=\frac{1}{4} a^2 \sqrt{2V(\log b^2)}\quad .
\]
These BPS topological defects are magnetic domain walls, the magnetic field in the $x_1$-direction is infinitely reproduced all over the $x_2:x_3$-plane. The BPS domain walls only exist in the south chart but there are also anti-kinks going down from the vacuum parallel to the south pole with asymptotic conditions:
\[
u(-\infty)=\log b^2 \, , \, \, \frac{du}{dx_1}(-\infty)=0 \, \, \, ; \, \, \, u(\infty)=0 \, , \, \, \frac{du}{dx_1}(\infty)=\sqrt{2V(\log b^2)}=2A_2(\infty) \, \, .
\]
Because there is invariance under the global phase transformation $f\to f\cdot e^{i\alpha}$, the corresponding kink/anti-kink orbits join the south pole with the vacuum parallel through all the meridians. There exists a degenerate in energy per unit surface family of magnetic domain walls parametrized by an angle similar to the family of kinks found in the supersymmetric $(1+1)$-dimensional $\mathbb{CP}^1$-sigma model in Reference \cite{Nitta} and further analyzed together with other BPS supersymmetric states in \cite{Nitta1,Nitta2}.

Finally, we mention in this context that there are also non-topological kink solutions to equation (\ref{mecqdw}) where the particle starts at $u(-\infty)=-\infty$, bounces back at the turning point $u(x_1^0)<\log b^2$ reached with zero velocity, $\frac{du}{dx_1}(x_1^0)=0$, and approaches again $u(+\infty)=-\infty$ in the remote future. Needless to say the corresponding magnetic walls are unstable.

\subsubsection{BPS scalar domain walls}

This structure of vacuum moduli space admits another kind of domain wall topological defects. The pertinent ansatz is:
\[
A_\mu(x^0,x^1,x^2,x^3)=0 \quad , \quad \xi(x^0,x^1,x^2,x^3)=f(x)  \quad ,
\]
where $f^*=f$ without loss of generality because of the global phase transformation. The electromagnetic field for all these configurations is zero and the scalar field is restricted to be real and varies only in the $x^1$-axis direction,  any spatial straight line passing through the origin in $\mathbb{R}^3$. The metric and the square root of the potential energy density
in the south chart read for these configurations:
\[
g_S(f^2)=\frac{4}{(f^2(x)+1)^2} \quad , \quad W_S(f^2)=\frac{\lambda}{2}\cdot\frac{a^2}{b^2}\cdot\frac{f(b^2-f^2)}{(f^2+1)\sqrt{f^2+d^2}}\quad .
\]
All this allows for a Bogomolny splitting of the energy per unit surface of a normalizing square of area $A$ in the orthogonal plane to the $x$-axis:
\begin{equation}
E/A=\frac{\rho^2}{2}\int dx \, g_S(f^2)\Big[\frac{df}{dx}-\frac{W_S(f^2)}{\rho\sqrt{g_S(f^2)}}\Big]^2 +\rho\int df\, \sqrt{g_S(f^2)} \, W_S(f^2)
\label{sdwfoe}\quad .
\end{equation}
Elementary integration of the BPS ODE coming from annihilation of the quadratic term in (\ref{sdwfoe})
\[
b^2\int \, df \, \frac{\sqrt{f^2+d^2}}{f(b^2-f^2)}=\mu (x-x_0) \quad \mbox{with} \quad \mu=\frac{\lambda}{4\rho}a^2
\]
leads to the expression
\begin{equation}
\sqrt{b^2+d^2}\, {\rm arctanh}\left[\sqrt{\frac{f^2+d^2}{b^2+d^2}}\right] - d \,  {\rm arctanh}\left[\frac{\sqrt{f^2+d^2}}{d}\right]=\mu(x-x_0) \label{bpssdw}
\end{equation}
which provides all the scalar domain wall trajectories by means of the implicit equation (\ref{sdwfoe}), see Figure 14. Taking the appropriate limits one checks that the kink profile starts at the south pole $f=0$ at $x=-\infty$ and ends at $x=+\infty$ in a point on the vacuum parallel $f=b$. Scalar anti domain walls are easily obtained changing $W_S$ by $-W_S$, the trajectories in this case run from the parallel to the south pole.

\begin{figure}[ht]
\centerline{\includegraphics[height=2.2cm]{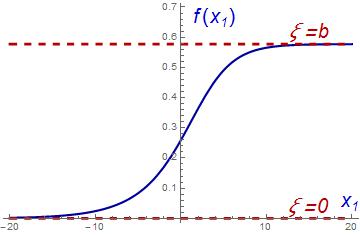}\hspace{1cm}
\includegraphics[height=2.2cm]{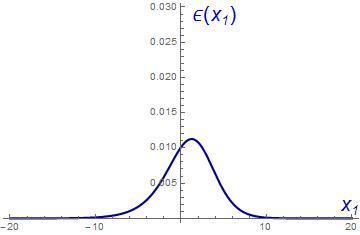}}
\caption{{\small Profiles of the BPS scalar domain walls for the case C ($b_1=0$ and $b_0=c_0=c_1=1$ with $e=\rho=a=1$): from left to right the function $f(x_1)$ and the energy density $\epsilon(x_1)$ per area unit.}}
\end{figure}

Integration of the second term in (\ref{sdwfoe})
\begin{eqnarray*}
K(f)&=&\rho\lambda \cdot\frac{a^2}{b^2}\cdot \int_0^b \, df \, \frac{f(b^2-f^2)}{(f^2+1)^2\sqrt{f^2+d^2}}=\\
&=&\frac{\rho\lambda}{2(1-d^2)} \cdot\frac{a^2}{b^2}\Big[(b^2+1)\frac{\sqrt{f^2+b^2}}{f^2+2}+\frac{2 d^2+b^2-1}{\sqrt{1-d^2}}{\rm arctan}\sqrt{\frac{f^2+d^2}{1-d^2}}\, \, \Big]
\end{eqnarray*}
unveils the finite surface tension of these BPS topological scalar domain walls in the analytical form: $
E^{BPS}/A=| K(b)-K(0)|$. We remark that there exist also a family of this kind of domain walls
due to invariance with respect to global phase transformation. Each BPS orbit runs along a meridian between the south pole
and the vacuum parallel.

\section{BPS topological defects in the Chern-Simons non-linear $\mathbb{CP}^1$-sigma model}

\subsection{Generalized Chern-Simons-Higgs Lagrangian \\ and the Bogomolny splitting}

Replacing Maxwell by Chern-Simons Abelian gauge theories is possible, and sensible, in $(2+1)$ dimensional space-time. The action governing the dynamics of these planar gauge theories is{\footnote{Because the dimension of the space-time is three instead four the physical dimensions of fields and parameters change: $[A_\mu]=[\varphi]=[\rho]=[a]=L^{-\frac{1}{2}}$. $[\kappa]=[e^2]=L^{-1}$.}}:
\begin{equation}
S_{\rm CSH}[A_\mu,\varphi]=\int \, d^3x \, \Big[\frac{\kappa}{4}\varepsilon^{\mu\nu\rho} A_\mu F_{\nu\rho}+\frac{1}{2}g(\vert\varphi\vert^2)D_\mu\varphi^*D^\mu\varphi-\frac{1}{2}W^2(\vert\varphi\vert^2)\Big] \label{gcsha} \, .
\end{equation}
Here, $\mu,\nu,\rho=0,1,2$ and $\varepsilon^{\mu\nu\rho}$ is the totally antisymmetric tensor. The usual Maxwell term is replaced in the Lagrangian by the secondary Chern-Simons density, dominant at low frequencies, a function $g(\vert\varphi\vert^2)$ is allowed in the Higgs field kinetic term, and the choice of the Higgs potential energy density as a perfect square announces a self-dual structure of the system. Variations with respect to $A_\mu$ give rise to the field equations:
\begin{equation}
\kappa\varepsilon^{\mu\nu\rho}F_{\nu\rho}=ie g(\vert\varphi\vert^2)\left[\varphi^*\cdot D^\mu\varphi-D^\mu\varphi^*\cdot\varphi\right] \label{cshfe}\, .
\end{equation}
The $\mu=0$ equation in (\ref{cshfe}), the Gauss law, allows to solve the $A_0$ potential in terms of the magnetic field and the electric charge density. For static fields one obtains:
\begin{equation}
A_0(x_1,x_2)=\frac{\kappa}{e^2\vert\varphi\vert^2}\cdot g^{-1}(\vert\varphi\vert^2)\cdot F_{12}(x_1,x_2) \label{gaussl}\quad .
\end{equation}
Plugging the constraint solution (\ref{gaussl}) in the expression for the static part of the energy coming from the action (\ref{gcsha}) we end with the formula:
\begin{equation}
E[A_j,\varphi]=\frac{1}{2}\int\, d^2x\, \Big[\frac{\kappa^2}{e^2\vert\varphi\vert^2}\cdot g^{-1}(\vert\varphi\vert^2)\cdot F_{12}^2+g(\vert\varphi\vert^2)\cdot D_j\varphi^*D_j\varphi+W^2(\vert\varphi\vert^2)\Big]\label{sencsh}) \, \, ,
\end{equation}
i.e., a self-dual system with dielectric function $H(\vert\varphi\vert^2)=\frac{\kappa^2}{e^2\vert\varphi\vert^2}\cdot g^{-1}(\vert\varphi\vert^2)$
provided that the potential energy density is chosen to be of the form:
\[
 U(\vert\varphi\vert^2)=\frac{1}{2}W^2(\vert\varphi\vert^2)=\frac{e^2}{8}\cdot \frac{\left(F(\vert\varphi\vert^2)-a^2\right)^2}{H(\vert\varphi\vert^2)}\, \, \, , \, \, \, F(\vert\varphi\vert^2)=\int \, g(\vert\varphi\vert^2) d \vert\varphi\vert^2 \, \, , \, \, F(0)=0 \, \, .
\]
In this case the regular solutions of the self-dual equations
\begin{equation}
B(x_1,x_2)=\pm \frac{e}{2}\cdot\frac{a^2-F(\vert\varphi\vert^2)}{H(\vert\varphi\vert^2)} \quad , \quad D_1\varphi(x_1,x_2)\pm i D_2\varphi(x_1,x_2)=0 \label{sdcsh}
\end{equation}
saturate the Bogomolny bound: $E_B=\frac{e}{2}a^2\int\, d^2x\, F_{12}=\pi\vert n\vert a^2$.

\subsection{Two species of BPS solitons in the Chern-Simons-Higgs gauged non-linear $\mathbb{S}^2$-sigma model}

In this subsection we address the promotion of the Jackiw-Weinberg Chern-Simons-Higgs theory \cite{Jackiw}, see also \cite{Fuerguil1} and the monography \cite{Dunne}, to a gauged nonlinear $\mathbb{S}^2$-sigma model.

\subsubsection{BPS topological vortices in the south chart}

We next consider the Chern-Simons-Higgs model when the scalar field takes values in the $\mathbb{S}^2$ target manifold, i.e., the functions described above are in the south chart:
\begin{eqnarray*}
&& H_S(\vert\phi\vert^2)=\frac{\kappa^2}{e^2\vert\phi\vert^2}\cdot\frac{(\rho^2+\vert\phi\vert^2)^2}{4\rho^4} \, \, \quad , \quad \, \, g_S(\vert\phi\vert^2)=
\frac{4\rho^4}{(\rho^2+\vert\phi\vert^2)^2} \\ && F_S(\vert\phi\vert^2)=\frac{4\rho^2\vert\phi\vert^2}{\rho^2+\vert\phi\vert^2}\, \, \, , \, \, \, \quad  U_S(\vert\phi\vert^2)=\frac{e^4}{2\kappa^2}\cdot\frac{\rho^4\vert\phi\vert^2}{(\rho^2+\vert\phi\vert^2)^2}\cdot\Big(\frac{4\rho^2\vert\phi\vert^2}{\rho^2+\vert\phi\vert^2}-a^2\Big)^2\, .
\end{eqnarray*}
The vacuum orbit in this system, the set of zeroes of $U_S(\vert\phi\vert^2)$, is the union of the south pole $\phi_0=0$, the parallel $\vert\phi_a\vert^2=\frac{a^2\rho^2}{4\rho^2-a^2}$, and the north pole $\phi_\infty=+\infty$:
\[
{\cal V}_S^{CS}= \Big\{v_0=0 \Big\} \, \bigsqcup \, \Big\{ \vert v_a\vert^2=\frac{a^2\rho^2}{4\rho^2-a^2}\,  \Big\}\bigsqcup \,\Big\{ v_\infty=+\infty \Big\} \, \, .
\]

\begin{figure}[ht]
\centerline{\includegraphics[height=2.5cm]{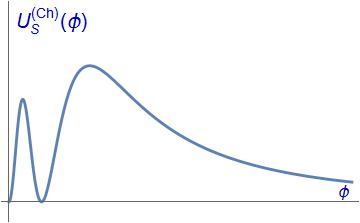}\hspace{2cm}
\includegraphics[height=2.5cm]{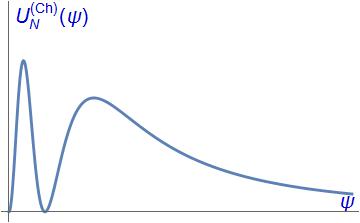}\hspace{0.3cm}}
\caption{\small Scalar potential for the CSH model on the $\phi$-chart (left) and $\psi$-chart (right) for $\kappa=a=\rho=e=1$.}
\end{figure}

The moduli space of vacua contains three points: ${\cal M}^{CS}={\cal V}_S^{CS}/\mathbb{U}(1)=\{v_0,v_a,v_\infty\}$, i.e., the south pole, any point in the vacuum parallel, and the north pole (see Figure 15).
There are two kinds of phases in this system. The particle spectrum in the symmetric phase, built either on $v_0$ or $v_\infty$, as ground state is formed by two charged scalar particles with
masses: $m_\phi^2=\frac{e^4}{4\kappa^2}a^4$. In the asymmetric phase where the ground state is one of the points in the vacuum parallel there is
a vector particle with only one polarization and a Higgs particle degenerate in mass:
\[
m_A^2=m_H^2=\frac{e^4}{4\kappa^2}\frac{(4\rho^2-a^2)^2}{4\rho^4}a^4 \quad .
\]
Regarding the self-dual structure it is clear that the first-order equations take the form:
\begin{equation}
B=\pm \frac{2 e^3 \rho^4\vert\phi\vert^2}{\kappa^2(\rho^2+\vert\phi\vert^2)^2}\cdot\left(a^2-\frac{4\rho^2\vert\phi\vert^2}{\rho^2+\vert\phi\vert^2}\right) \quad , \quad D_1\phi\pm i D_2\phi=0 \label{sdcshnlsm} \, \, \, .
\end{equation}
The functions $f(r)$ and $\beta(r)$, which characterize the radially symmetric solutions of (\ref{sdcshnlsm}) in this new context are drawn in Figure 16 together its energy density and magnetic field. The found pattern is similar to the solutions which arose in the models with a dielectric function with the parameter $b_1=0$, see Figure 12.

\begin{figure}[ht]
\centerline{\includegraphics[height=2.4cm]{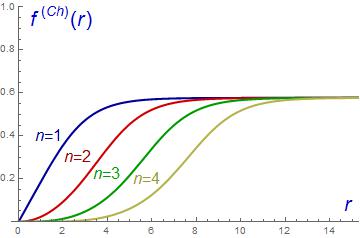}\hspace{0.1cm}
\includegraphics[height=2.4cm]{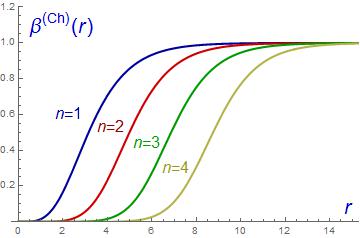}\hspace{0.1cm}
\includegraphics[height=2.4cm]{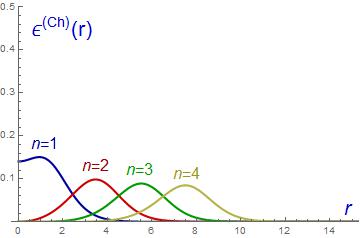}\hspace{0.1cm}
\includegraphics[height=2.4cm]{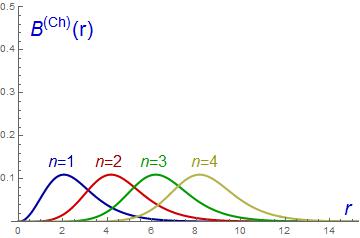}}
\caption{{\small Profiles of the radially symmetric $\phi$-vortices for several values of the vorticity $n$ in the CSH model with $\kappa=a=\rho=1$: from left to right the function $f(r)$, the function $\beta(r)$, the energy density $\epsilon(r)$ and the magnetic field $B(r)$.}}
\end{figure}

\subsubsection{BPS topological vortices in the north chart}
In the north chart the dielectric function becomes:
\[
H_N(\vert\psi\vert^2)=H_S \Big( \frac{\rho^4}{\vert\psi\vert^2} \Big) =\frac{\kappa^2}{e^2\vert\psi\vert^2}\cdot\frac{(\rho^2+\vert\psi\vert^2)^2}{4\rho^4}\, \, \, .
\]
Moreover,
\[
g_N(\vert\psi\vert^2)D_\mu\psi^*D^\mu\psi=g_S \Big( \frac{\rho^4}{\vert\psi\vert^2} \Big) D_\mu\left(\frac{\rho^2}{\psi}\right)D^\mu\left(\frac{\rho^2}{\psi^*}\right), \, \, \, 4\rho^2-F_N(\vert\psi\vert^2)= F_S \Big(\frac{\rho^4}{\vert\psi\vert^2} \Big)
\]
implies that:
\begin{eqnarray*}
&& g_N(\vert\psi\vert^2)=\frac{4\rho^4}{(\rho^2+\vert\psi\vert^2)^2} \, \, , \hspace{1cm}\, \, \, F_N(\vert\psi\vert^2)=\frac{4\rho^2\vert\psi\vert^2}{\rho^2+\vert\psi\vert^2}\\ && U_N(\vert\psi\vert^2)=\frac{e^4}{2\kappa^2}\cdot\frac{\rho^4\vert\psi\vert^2}{(\rho^2+\vert\psi\vert^2)^2}\cdot \Big(\frac{4\rho^2\vert\psi\vert^2}{\rho^2+\vert\psi\vert^2}-4\rho^2+a^2\Big)^2 \quad .
\end{eqnarray*}
Therefore, like in the south chart the vacuum orbit is the disjoint union of the north pole $\psi_0=0$,the vacuum parallel $\vert\psi_a\vert^2=\rho^2\frac{4\rho^2-a^2}{a^2}$, and the south pole $\psi_\infty=+\infty$:
\[
{\cal V}_N^{CS}= \Big\{ w_0=0 \Big\} \, \bigsqcup \, \Big\{ \vert w_a\vert^2=\rho^2\frac{4\rho^2-a^2}{a^2} \Big\}\, \bigsqcup \Big\{ \, w_\infty=+\infty \Big\} \, \, .
\]
The moduli space of vacua is again formed by the two poles as isolated zeroes of $U_N$ and the vacuum parallel, see Figure 15. In the symmetric phases there is a pair of charged bosons and masses: $m_\psi^2=\frac{e^4}{4\kappa^2}\cdot(4\rho^2-a^2)^2$. A vector boson with only one polarization and a Higgs particle with masses
\[
m_A^2=m_H^2=\frac{e^4}{4\kappa^2}\cdot\frac{(4\rho^2-a^2)^2}{4\rho^4}\cdot a^2
\]
arises in the asymmetric phase. The system is globally well defined.
The first-order Bogomolny equations read in the north chart:
\begin{equation}
B=\pm \frac{2 e^3 \rho^4\vert\psi\vert^2}{\kappa^2(\rho^2+\vert\psi\vert^2)^2}\cdot\left(4\rho^2-a^2-\frac{4\rho^2\vert\psi\vert^2}{\rho^2+\vert\psi\vert^2}\right) \quad , \quad D_1\psi\pm i D_2\psi=0 \label{sdcshnlsm1} \, \, \, .
\end{equation}
For sake of completeness Figure 17 shows the behaviour of the radially symmetric $\psi$-vortices. Notice that for our choice of the parameters the $\psi$-vortices exhibit a more concentrated energy density and magnetic fields than the $\phi$-vortices, as we can observe by comparing Figures 16 and 17.

\begin{figure}[ht]
\centerline{\includegraphics[height=2.4cm]{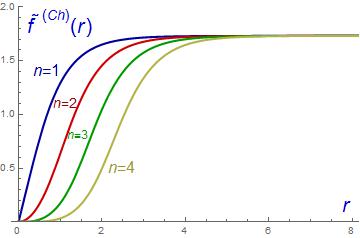}\hspace{0.1cm}
\includegraphics[height=2.4cm]{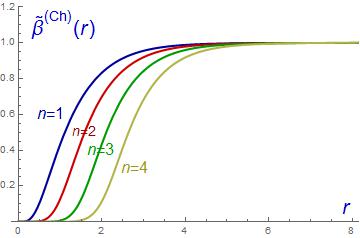}\hspace{0.1cm}
\includegraphics[height=2.4cm]{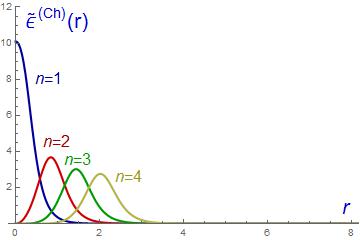}\hspace{0.1cm}
\includegraphics[height=2.4cm]{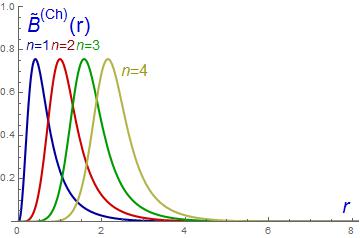}}
\caption{{\small Profiles of the radially symmetric $\psi$-vortices for several values of the vorticity $n$ in the CSH model with $\kappa=a=\rho=1$: from left to right the function $f(r)$, the function $\beta(r)$, the energy density $\epsilon(r)$ and the magnetic field $B(r)$.}}
\end{figure}

\subsubsection{BPS magnetic domain ribbons}
Coming back to the south chart we search for BPS magnetic domain ribbons trying the ansatz:
\[
A_1=0 \quad , \quad A_2=A_2(x_1) \quad , \quad \phi=\rho f(x_1) \quad , \quad f^*=f \quad ,
\]
i.e., relying on the axial gauge $A_1=0$ we focus on fields depending only on the $x_1$-coordinate and such that the scalar field is real.
The BPS defects with these properties are magnetic ribbons, constant along the $x_2$-direction, forming a boundary between two domains in the plane. BPS domain ribbons along the $x_1$-axis would be obtained from a similar ansatz exchanging the sub-indices $1$ and $2$. For this ansatz
the self-dual equations (\ref{sdcshnlsm}) become:
\[
\frac{d A_2}{dx_1}=\pm 2\frac{e^3\rho^2}{\kappa^2}\frac{a^2}{b^2}\frac{f^2(b^2-f^2)}{(1+f^2)^3} \, \, \, \,  , \quad \, \, A_2(x_1)=\mp \frac{1}{e}\frac{d}{dx_1}(\log f(x_1)) \, \, .
\]
where $b=\frac{a}{\sqrt{4 \rho^2-a^2}}$. Denoting $f(x_1)=e^{u(x_1)/2}$ this ODE system is rewritten as the second-order ODE:
\begin{equation}
\frac{d^2u}{dx_1^2}=-2 C \frac{e^u(b^2-e^u)}{(1+e^u)^3}=-\frac{dV}{du}(u)\, \, \, , \, \, \, C=\frac{2 e^4\rho^2}{\kappa^2}\frac{a^2}{b^2}>0 \label{medrsc}\quad .
\end{equation}
This is a  mechanical problem with potential energy
\[
V(u)=C \frac{2 b^2+(b^2-1)e^u}{4 \cosh^2(\frac{u}{2})} \quad,
\]
where the integration constant has been set such that $V(-\infty)=0$. This potential energy has a maximum at $u=\log b^2$ where $V(\log b^2)=\frac{b^4}{1+b^2}C$ and reaches the asymptotic value $ V=(b^2-1)C$ for $u\rightarrow +\infty$. For a trajectory starting at the south pole and and ending at the vacuum parallel, i.e. such that
\bdm
 u(-\infty)=-\infty,\quad \quad u(\infty)=\log b^2,\quad \quad  \frac{du}{dx_1}(\infty)=0
\edm
the initial speed is $\frac{du}{dx_1}(-\infty)=\sqrt{2V(\log b^2)}$ and the energy first-integral $I=\frac{1}{2}(\frac{du}{dx_1})^2+V(u)$ takes the value $I=V(\log b^2)$. An elementary integration allows the description of these trajectories in terms of the implicit equation:
\begin{equation}
u-(1+b^2) \log\left(b^2-e^u\right)=\frac{e^2 a^2}{\kappa}(x_1-x_1^0) \quad . \label{cshdrs}
\end{equation}
Of course, exchanging the sign of the right hand side of formula (\ref{cshdrs}) one obtains the BPS antikinks and, like in previous cases of domain walls, there are kink and antikinks orbits along all the meridians due to phase invariance of the system. In this $(2+1)$-dimensional space-time these BPS defects are reproduced infinitely along the $x_2$-axis. Thus, they are thick strings or domain ribbons with an string tension or energy per unit length
\beq
E/L=\frac{1}{2}e a^2 \left[A_2(+\infty)-A_2(-\infty)\right]=\frac{1}{4} a^2 \sqrt{2V(\log b^2)}=\frac{e^2 a^4}{4\kappa} \label{tdrs}.
\eeq
Apart from these solutions, there are also trajectories with initial conditions
\bdm
 u(-\infty)=-\infty, \quad \quad \frac{du}{dx_1}(-\infty)<\sqrt{2V(\log b^2)}
\edm
which have a turning point for $u_{MAX}<\log b^2$. These are the south-chart non-topological ribbons which interpolate between two domains where the fields have settled down at the south pole vacuum. They have string tension
\[
E/L=\frac{1}{2} a^2 \sqrt{2V(u_{MAX})}
\]
and, in the limit $u_{MAX}\rightarrow \log b^2$, they can be thought of as composite systems containing a topological ribbon and antiribbon.

Since the dielectric function and potential energy density have the same structure both the north and south charts, the analysis leading to the north-chart magnetic ribbons is carried out in an entirely analogous fashion. The only relevant change is $a^2\rightarrow 4\rho^2-a^2$ for the vacuum condensate, which implies also $b\rightarrow \frac{1}{b}$. The topological ribbons and antiribbons making the transition between the north pole and the vacuum parallel are thus given by the implicit equation
\bdm
\tilde{u}-\frac{1+b^2}{b^2} \log\left(1-b^2 e^{\tilde{u}}\right)=\pm\frac{e^2 (4\rho^2- a^2)}{\kappa}(x_1-x_1^0), \quad \quad \psi(x_1)=\rho e^\frac{\tilde{u}(x_1)}{2}
\edm
They are depicted in Figure 18 and 19 together its energy densities and magnetic fields. The string tension is
\[
E/L=\frac{e^2 (4\rho^2-a^2)^2}{4\kappa},
\]
while there are also non-topological ribbons starting and ending at the north pole which are the twins of the antipodal ones in the south chart.
\begin{figure}[ht]
\centerline{\includegraphics[height=2.2cm]{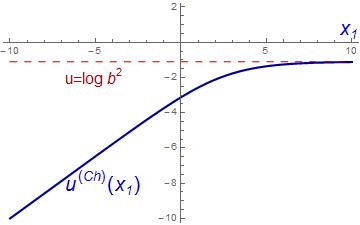}\hspace{0.1cm}
\includegraphics[height=2.2cm]{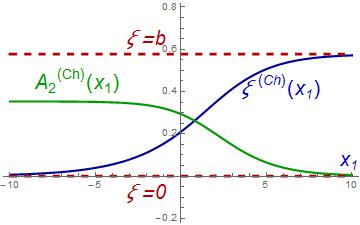}\hspace{0.1cm}
\includegraphics[height=2.2cm]{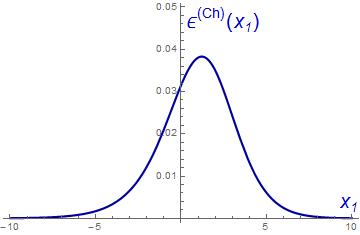}\hspace{0.1cm}
\includegraphics[height=2.2cm]{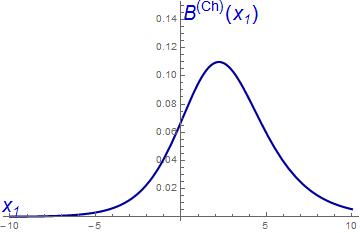}}
\caption{{\small Profiles of the BPS magnetic $\phi$ domain walls (in the south chart) for the case CSH model with $\kappa=a=\rho=1$: from left to right the function $u(x_1)$, the functions $\xi(x_1)$ and $A_2(x_1)$, the energy density $\epsilon(x_1)$ and the magnetic field $B(x_1)$ per length unit.}}
\end{figure}
\begin{figure}[ht]
\centerline{\includegraphics[height=2.2cm]{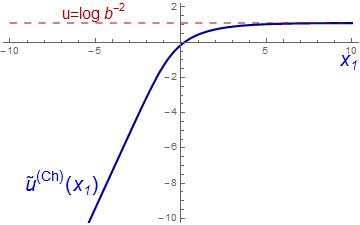}\hspace{0.1cm}
\includegraphics[height=2.2cm]{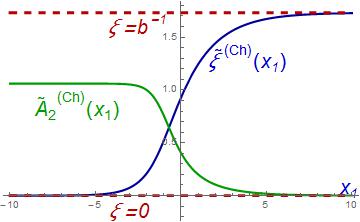}\hspace{0.1cm}
\includegraphics[height=2.2cm]{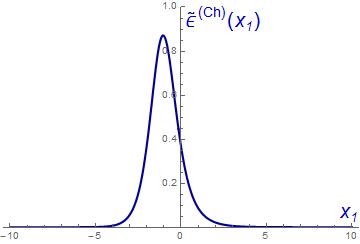}\hspace{0.1cm}
\includegraphics[height=2.2cm]{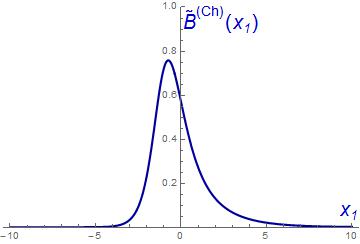}}
\caption{{\small Profiles of the BPS magnetic $\psi$ domain walls (in the north chart) for the case CSH model with $\kappa=a=\rho=1$: from left to right the function $\widetilde{u}(x_1)$, the functions $\widetilde{\xi}(x_1)$ and $\widetilde{A}_2(x_1)$, the energy density $\widetilde{\epsilon}(x_1)$ and the magnetic field $\widetilde{B}(x_1)$ per length unit.}}
\end{figure}

\subsubsection{BPS scalar domain ribbons}
We now search for purely scalar domain ribbon BPS solutions depending only on a coordinate $x$ along an arbitrary axis in the $x_1:x_2$-plane. The metric and the square root of the potential energy density
in the south chart read for these configurations:
\[
g_S(f^2)=\frac{4}{(f^2(x)+1)^2} \quad , \quad W_S(f^2)=\frac{e^2 \rho a^2}{\kappa b^2}\cdot\frac{f(b^2-f^2)}{(f^2+1)^2}\quad .
\]
The scalar domain ribbons obey the BPS equation
\bdm
\frac{df}{dx}=\frac{W_S(f^2)}{\rho\sqrt{g_s(f^2)}}=\frac{e^2 a^2}{2 \kappa b^2} \frac{f(b^2-f^2)}{1+f^2}
\edm
are thus determined from the implicit equation:
\begin{equation}
\log f^2-(1+b^2)\log (b^2-f^2)=\pm\frac{e^2 a^2}{\kappa}\cdot (x-x_0) .\label{cshesc}
\end{equation}
These BPS kinks give rise to scalar domain ribbon orbits running from the south pole to the vacuum parallel or from that parallel to the south pole depending on the sign, see Figure 20. Because the phase invariance $f\to f\cdot e^{i\alpha}$ there is a domain ribbon orbit on each meridian. The energy per unit length of these solitons is $E/L= K(b)-K(0)$ where
\[
K(f)=\rho\int \, df\, \sqrt{g_s(f^2)}W_S(f^2)=\frac{2 e^2 \rho^2 a^2}{\kappa b^2} \int\, df \, \frac{f(b^2-f^2)}{(f^2+1)^3}
=\frac{e^2 \rho^2 a^2}{2 \kappa b^2} \cdot\frac{1-b^2+2 f^2}{(1+f^2)^2}
\]
and gives the result
\beq
E/L=\frac{e^2 a^4}{8\kappa} \quad .\label{tesc}
\eeq
Comparison of the trajectory (\ref{cshdrs}) with the solution (\ref{cshesc}) shows a curious fact: in the Chern-Simons-Higgs model the scalar field profile is the same for magnetic and scalar domain ribbons. Nevertheless, the resulting string tensions (\ref{tdrs}) and (\ref{tesc}) are different, because in the former case there are contributions to the energy coming both from the scalar and vector components of the solution. As a consequence of the BPS equations, the two contributions are exactly the same and the string tension of the magnetic domain ribbon is thus twice that of the purely scalar defect.

\begin{figure}[ht]
\centerline{\includegraphics[height=2.2cm]{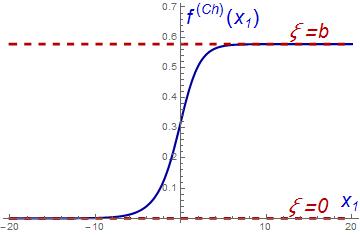}\hspace{0.1cm}
\includegraphics[height=2.2cm]{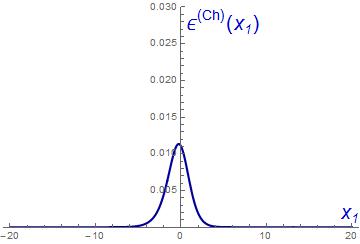}}
\caption{{\small Profiles of the BPS scalar $\phi$ domain walls (in the south chart) for the CSH model with $\kappa=\rho=a=1$): from left to right the function $f(x_1)$ and the energy density $\epsilon(x_1)$ per length unit.}}
\end{figure}

\begin{figure}[ht]
\centerline{\includegraphics[height=2.2cm]{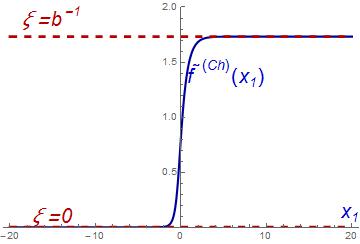}\hspace{0.1cm}
\includegraphics[height=2.2cm]{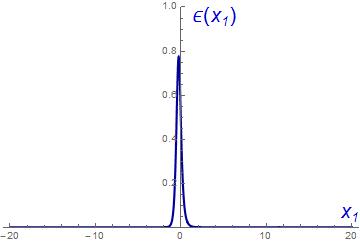}}
\caption{{\small Profiles of the BPS scalar $\psi$ domain walls (in the north chart) for the CSH model with $\kappa=\rho=a=1$): from left to right the function $\widetilde{f}(x_1)$ and the energy density $\widetilde{\epsilon}(x_1)$ per length unit.}}
\end{figure}

The scalar domain ribbons running from the north between the pole and the vacuum parallel come, as before, through the changes $a^2\rightarrow 4\rho^2-a^2$, $b\rightarrow \frac{1}{b}$ in the previous expressions, see Figure 21. In particular, the string tension of these defects is
\bdm
E/L=\frac{e^2 (4\rho^2- a^2)^2}{8\kappa} \quad.
\edm

\section{BPS solitons in the massive gauged non-linear $\mathbb{C P}^2$-sigma model}

The non-linear sigma model on the sphere is the simplest $n=1$ representative among a hierarchy of $\mathbb{U}(1)$-gauged non-linear sigma models with $\mathbb{C P}^n$ manifolds as target spaces. Models of this class arise quite naturally as effective theories on the world sheet of nonabelian strings, see for instance \cite{shifm,shyu} and references therein. In order to extend the results found in $\mathbb{S}^2\simeq\mathbb{CP}^1$ to other members of the hierarchy, in this section  we shall analyse the next case, since it turns out that $\mathbb{CP}^2$ already exhibits the most relevant features arising for general $n$. $\mathbb{CP}^2$ is a K\"{a}hler manifold of complex dimension two. A coordinate system is built from a minimal atlas with three charts. We shall call them $V_\phi, V_\psi$ and $V_\xi$ and shall denote the complex coordinates in each chart, respectively, as $(\phi_1,\phi_2)$, $(\psi_1,\psi_2)$ and $(\xi_1,\xi_2)$. The transition functions giving the change of coordinates in the intersections between charts are: $\psi_1=\frac{\rho^2}{\phi_1}, \psi_2=\rho\frac{\phi_2}{\phi_1}$ on $V_\phi\cap V_\psi$, $\xi_1=\rho\frac{\phi_1}{\phi_2}, \xi_2=\frac{\rho^2}{\phi_2}$ on $V_\phi\cap V_\xi$ and $\xi_1=\frac{\rho^2}{\psi_2}, \xi_2=\rho\frac{\psi_1}{\psi_2}$ on $V_\psi\cap V_\xi$. The K\"{a}hler potential,  expressed in the  coordinates of the chart $V_\phi$, takes the form $K=4\rho^2\ln (1+\frac{|\phi_1|^2}{\rho^2}+\frac{|\phi_2|^2}{\rho^2})$ and it is easy to check that it adopts an equivalent form, in terms of the respective coordinates, on the other two charts.

\subsection{The gauged Abelian $\mathbb{C P}^2$-sigma model in the reference chart $V_{\phi}$}
To formulate an Abelian gauge theory on $\mathbb{C}\mathbb{P}^2$ let us begin by working on the $V_\phi$ chart. Gauging of the scalar non-linear $\mathbb{C}\mathbb{P}^2$ sigma model gives rise to the Lagrangian
\beq
{\cal L}_\phi=-\frac{1}{4} F_{\mu\nu} F^{\mu\nu} +\frac{1}{2} g_{p\bar{q}} D_\mu \phi_p D^\mu \phi_q^*-U_\phi(|\phi_1|^2,|\phi_2|^2)\ \label{lag1} ,
\eeq
where $g_{p\bar{q}}=\frac{\partial^2 K}{\partial\phi_p\partial\phi_q^*}$ is the standard Fubini-Study metric on $\mathbb{C}\mathbb{P}^2$:
\begin{eqnarray*}
g_{1\bar{1}}&=&\frac{4\rho^2(\rho^2+\vert\phi_2\vert^2)}{(\rho^2+\vert\phi_1\vert^2+\vert\phi_2\vert^2)^2} \quad , \quad g_{1\bar{2}}=-\frac{4\rho^2 \phi_1^*\phi_2}{(\rho^2+\vert\phi_1\vert^2+\vert\phi_2\vert^2)^2} \\ g_{2\bar{2}}&=&\frac{4\rho^2(\rho^2+\vert\phi_1\vert^2)}{(\rho^2+\vert\phi_1\vert^2+\vert\phi_2\vert^2)^2} \quad , \quad g_{\bar{1}2}=-\frac{4\rho^2 \phi_1\phi_2^*}{(\rho^2+\vert\phi_1\vert^2+\vert\phi_2\vert^2)^2} \quad .
\end{eqnarray*}
The covariant derivatives are $D_\mu\phi_p=\partial_\mu\phi_p-i e A_\mu\phi_p$ and, to keep things simple, here we have discarded the possibility of introducing a dielectric function. We choose the potential energy density in (\ref{lag1}) in the form that generalizes the potential function entering the $\mathbb{S}^2$-sigma model.
\beq
U_\phi=\frac{e^2}{8}\left(\frac{4\rho^2(|\phi_1|^2+|\phi_2|^2)}{\rho^2+|\phi_1|^2+|\phi_2|^2}-a^2\right)^2 \quad .
\eeq
The Lagrangian corresponds to a semi-local theory  in which the $\mathbb{U}(1)$ local invariance is accompanied by a global $\mathbb{S U}(2)$ symmetry. The vacuum orbit, the set of zeroes of $U_\phi$, $\phi_1=v_1$, $\phi_2=v_2$, is the $\mathbb{S}^3$ sphere:
\[
\vert v_1\vert^2+|v_2|^2=\frac{\rho^2 a^2}{4\rho^2-a^2} \quad .
\]
It is well known that $\mathbb{S}^3$ is a Hopf bundle, i.e., it is a manifold fibered on $\mathbb{S}^2$ with fibre $\mathbb{S}^1$ and Hopf index $1$, see e.g. References \cite{AWJM0,AWJM}. Moreover, the winding number of the map from the $\mathbb{S}^1_\infty$ circle enclosing the spatial plane to the $\mathbb{S}^1$ fiber provided by the gauge field at infinity classifies the configuration space in $n\in\mathbb{Z}$ disconnected subspaces.
In the $n=0$ subspace, one sets a particular point of $\mathbb{S}^3$ as the vacuum, for instance $(v_1=\frac{\rho a}{\sqrt{4\rho^2-a^2}}, v_2=0)$. The Higgs mechanism is worked out and gives mass to the physical fields.  One thus finds that the mass spectrum includes, along with a degenerate couple encompassing the Higgs scalar meson and a massive vector boson, another complex Goldstone boson. The masses are:
\[
m_A^2=m_H^2=\frac{e^2}{4} a^2 \Big(1-\frac{a^2}{4\rho^2}\Big)^2=m^2 \quad , \quad m_G^2=0 \, \, .
\]
Both the Higgs and the Goldstone fields are coupled to the gauge field through the covariant derivative terms. All this refers to elementary quanta; the other perspective is about solitons living in sectors of the configuration space with $n\neq 0$.

In this respect, we look at the static part of the energy per unit length. Working in the simultaneous temporal and axial gauge, and focusing on static and $x_3$-independent configurations, one writes:
\beq
E=\int \, dx^2 \, \left\{\frac{1}{2} B^2+\frac{1}{2} g_{p\bar{q}} D_i \phi_p D_i \phi_q^*+\frac{e^2}{8}\Big(\frac{4\rho^2(\vert\phi_1\vert^2+\vert\phi_2\vert^2)}{\rho^2+\vert\phi_1\vert^2+\vert\phi_2\vert^2}-a^2\Big)^2\right\} \label{ener1}
\eeq
On one hand, we have that:
\begin{eqnarray*}
\frac{1}{2}\Big(B^2+W^2 \Big)&=&\frac{1}{2}\Big(B\pm W\Big)^2\mp BW\\
W&=& \frac{e}{2}\Big(\frac{4\rho^2(\vert\phi_1\vert^2+\vert\phi_2\vert^2)}{\rho^2+\vert\phi_1\vert^2+\vert\phi_2\vert^2}-a^2\Big) \quad .
\end{eqnarray*}
On the other hand, we also split the covariant derivative terms in a similar manner
\bdm
\frac{1}{2} g_{p\bar{q}} D_k \phi_p D_k \phi_q^*=\frac{1}{2} g_{p\bar{q}}(D_1\phi_p\mp i D_2\phi_p)(D_1\phi_q^* \pm i D_2\phi_q^*)\mp R
\edm
where the last term can be conveniently recast in the form:
\[
R=\frac{i}{2} \varepsilon_{ij} M_{pq} (\partial_i \ln |\phi_p|-ie V_{ip}) (\partial_j \ln |\phi_q|+ie V_{jq}) \quad , \quad V_{ip}=A_i-\frac{1}{e} \partial_i \alpha_p \quad .
\]
Here $\alpha_p$ is the phase of $\phi_p$ and $M_{pq}=g_{p\bar{q}} \phi_p\phi_q^*$ (no sum in p and q) is a real symmetric matrix. Upon discarding total derivative terms, we write $R=e\varepsilon_{ij} V_{ip} S_{jp}$, where
\[
S_{jp}=M_{pq}\partial_j \ln|\phi_p|=2\rho^2\partial_j\frac{ |\phi_p|^2}{\rho^2+|\phi_1|^2+|\phi_2|^2} \quad ,
\]
and we find as the final outcome of all these manipulations that:
\beq
 R=2 e \rho^2\frac{|\phi_1|^2+|\phi_2|^2}{\rho^2+|\phi_1|^2+|\phi_2|^2} B \quad .
 \eeq
Therefore, we finish with a self-dual theory on the chart $V_\phi$ where the Bogomolny bound
\beq
 E\geq \frac{1}{2} e a^2 |\Phi_M| \quad , \quad e \Phi_M= e \int  d^2x B =2\pi n
\eeq
is saturated if the Bogomolny equations
\beq
B=\pm\frac{e}{2}\left(a^2-\frac{4
\rho^2(|\phi_1|^2+|\phi_2|^2)}{\rho^2+|\phi_1|^2+|\phi_2|^2}\right)\hspace{1cm} D_1\phi_p\pm i D_2\phi_p=0, p=1,2 \label{slsdbog}
\eeq
are satisfied. As required by the mixing of $\mathbb{S U}(2)$ and $\mathbb{U}(1)$ symmetries, these first-order equations are of the semi-local type. Although the equation for the magnetic field is more complicated here than in the standard examples, the results found in references lsuch as \cite{vaachu, hind, gors} give a solid guarantee that semi-local vortices enjoying stability against decay to the vacuum and filling, for winding number $n$, a moduli space of complex dimension $2 n$, will also exist in the present situation.

\subsubsection{Radially symmetric semi-local topological solitons}
The radial ansatz for the fields
\begin{eqnarray*}
\phi_1(r)&=&f(r)e^{i n \theta} \, \, , n\in\mathbb{Z}^+ \,  \quad , \hspace{1cm} \, r A_\theta(r)=n\beta(r)
\\ \phi_2(r)&=&\vert h(r)\vert e^{i(\omega+l \theta)} \, \, , \quad  \, l=0,1, \cdots, n \, \, ,\, \quad \omega\in\mathbb{R}^+ \quad ,
\end{eqnarray*}
converts the PDE Bogomolny system (\ref{slsdbog}) in the first-order ODE system
\begin{eqnarray}
\frac{n}{r}\frac{d\beta}{dr}&=& \pm\frac{e}{2}\Big[a^2-\frac{4\rho^2(f^2(r)+\vert h(r)\vert^2)}{\rho^2+f^2(r)+\vert h(r)\vert^2}\Big] \label{radsl1}\\
\frac{df}{dr}&=&\frac{n}{r}f(r)[1-\beta(r)] \quad , \quad \frac{d\vert h\vert}{dr}=\frac{n}{r}\vert h(r)\vert \Big[\frac{l}{n}-\beta(r)\Big] \label{radsl2} \quad .
\end{eqnarray}
The solutions of this system (\ref{radsl1})-(\ref{radsl2}) complying with the asymptotic conditions
\begin{eqnarray*}
&& \lim_{r\to \infty}f(r)=v_1 \quad , \quad \lim_{r\to \infty}h(r)=0 \quad , \quad \lim_{r\to +\infty}\beta(r)=1 \\
&& f(0)=0 \quad \quad , \quad \quad \vert h(0)\vert=\vert h_0\vert \delta_{l,0} \qquad , \quad \quad \beta(0)=0
\end{eqnarray*}
are the BPS solitons of the system in the $V_\phi$ (reference) chart. The choice $h_0=0$ gives rise to the Nitta-Vinci vortices in the south chart of $\mathbb{S}^2$ embedded in the reference chart of $\mathbb{C P}^2$. Setting $h_0 > 0$, cousins of the planar semi-local topological solitons arise in the gauged $\mathbb{C P}^n$ model. In fact, in the limit $h_0\to +\infty$ $\mathbb{C P}^1$-lumps appear where the magnetic flux is homogeneously spread throughout the plane. The topology behind their existence lies in the $\mathbb{S}^2$ base space of the Hopf fibration. To illustrate these points,  in Figures 22 and 23 we show two self-dual solutions for $h_0=0.1$ and $h_0=0.4$ obtained by means of the same shooting procedure as applied before. In the first case, the soliton profiles, the energy density per unit length and the magnetic field are quite close to their counterparts in the Nitta-Vinci vortices. For higher $h_0$,
we see that the profiles of the solutions, together with the energy density per unit length and magnetic field, are less concentrated and tend slowly to their vacuum values.

\begin{figure}[ht]
\centerline{\includegraphics[height=2.5cm]{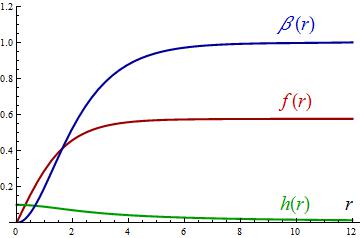}\hspace{0.6cm}
\includegraphics[height=2.5cm]{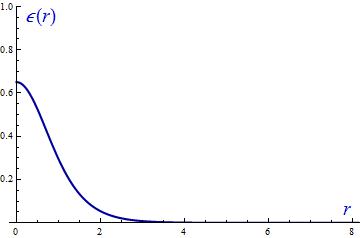}\hspace{0.6cm}
\includegraphics[height=2.5cm]{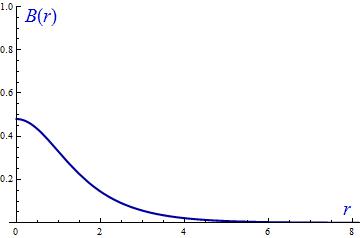}}
\caption{\small Profiles of the functions $f(r)$, $\beta(r)$ and $h(r)$ (left), the energy density (middle) and the magnetic field $B(r)$ for
vorticity $n=1$ and the choice $h_0=0.1$.}
\end{figure}

\begin{figure}[ht]
\centerline{\includegraphics[height=2.5cm]{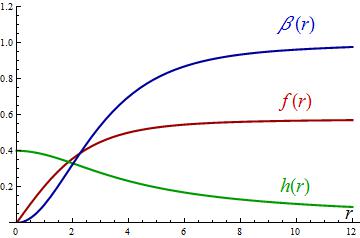}\hspace{0.6cm}
\includegraphics[height=2.5cm]{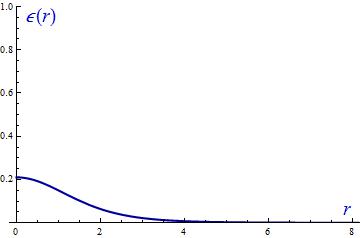}\hspace{0.6cm}
\includegraphics[height=2.5cm]{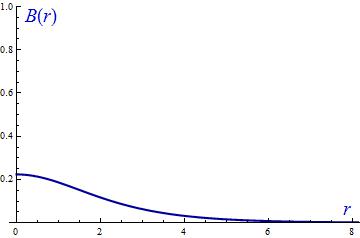}}
\caption{\small Profiles of the functions $f(r)$, $\beta(r)$ and $h(r)$ (left), the energy density (middle) and the magnetic field $B(r)$ for
vorticity $n=1$ and the choice $h_0=0.4$.}
\end{figure}

\subsection{The gauged Abelian $\mathbb{C P}^2$-sigma model in the second chart}
The next task is to write the Lagrangian (\ref{lag1}) as an Abelian theory in the chart $V_\psi$. In order to do so, we first observe that the K\"{a}hler potential
becomes
\[
K=4\rho^2\Big[ \ln \Big(1+\frac{\vert\psi_1\vert^2}{\rho^2}+
\frac{\vert\psi_2\vert^2}{\rho^2}\Big)-\ln\frac{\psi_1^*\psi_1}{\rho^2}\Big]
\]
in the new chart. Thus, the Fubini-Study metric remains formally identical. The changes in the covariant derivatives
\[
D_\mu\phi_1=-\frac{\rho^2}{\psi_1^2}(\partial_\mu\psi_1+i e A_\mu\psi_1) \, \, \, , \, \, \, D_\mu\phi_2=-\rho\frac{\psi_2}{\psi_1^2}(\partial_\mu\psi_1+i e A_\mu\psi_1) +\frac{\rho}{\psi_1}\partial_\mu\psi_2 \nonumber
\]
seem to be more drastic but are explained as follows: the gauge transformations $\phi_p\rightarrow e^{i e \chi} \phi_p, p=1,2$  of the $\phi$-fields correspond, after the application of the transition functions on $V_\phi\cap V_\psi$, to the local phase redefinitions $\psi_1\rightarrow e^{-i e \chi}\psi_1$ and $\psi_2\rightarrow \psi_2$ for the $\psi$-fields. The $\psi_1$ field couples to the gauge field with opposite charge to the two $\phi_p$ fields,  but $\psi_2$ remains as a neutral field. Consequently, the covariant derivatives on the chart $V_\psi$ are defined as:
\[
D_\mu\psi_p=\partial_\mu\psi_p-i N_p e A_\mu\psi_p \, \, , \hspace{1cm} {\rm where} \hspace{1cm} N_1=-1 \hspace{0.5cm}{\rm and} \hspace{0.5cm}N_2=0 \, \,
\]
and have the right holomorphic transformation properties under the change of chart:
\[
D_\mu\phi_p=\frac{\partial\phi_p}{\partial\psi_q} D_\mu\psi_q \, \,  ; \quad  \, \, \frac{\partial\phi_1}{\partial\psi_1}=-\frac{\rho^2}{\psi_1^2} \, \, \,  , \, \, \, \frac{\partial\phi_1}{\partial\psi_2}=0 \, \, \,  , \, \, \, \frac{\partial\phi_2}{\partial\psi_1}= -\rho\frac{\psi_2}{\psi_1^2} \, \, \, , \, \, \, \frac{\partial\phi_2}{\partial\psi_2}=\frac{\rho}{\psi_1} \, \, .
\]
This is all that is needed to ensure that the K\"{a}hler structure of the kinetic term is preserved, leading us to the Lagrangian in the chart $V_\psi$:
\beq
{\cal L}_\psi=-\frac{1}{4} F_{\mu\nu} F^{\mu\nu} +\frac{1}{2} g_{p\bar{q}} D_\mu \psi_p D^\mu \psi_q^*-U_\psi(|\psi_1|^2,|\psi_2|^2)\ \label{lag2} .
\eeq
$g_{p\bar{q}}$ is the Fubini-Study metric defined in terms of the $\psi$ fields, and the new potential energy density is
\beq
U_\psi=\frac{e^2}{8}\left(\frac{4\rho^2|\psi_1|^2}{\rho^2+|\psi_1|^2+|\psi_2|^2}-(4\rho^2-a^2)\right)^2. \label{pot2}
\eeq
Alternatively, one can derive the Lagrangian (\ref{lag2}) from the Lagrangian (\ref{lag1}) by applying the transition functions in a direct but lengthy calculation.

The potential energy density (\ref{pot2}) is only invariant under the Abelian subgroup
\[
\left(\begin{array}{c}\psi_1^\prime\\ \psi_2^\prime\end{array}\right)=\left(\begin{array}{cc}e^{i\alpha} & 0 \\0 & e^{-i\alpha}\end{array}\right)\left(\begin{array}{c}\psi_1 \\ \psi_2\end{array}\right)
\]
of the global $\mathbb{S U}(2)$ non-Abelian symmetry. Because $\psi_2$ is neutral the local $\mathbb{U}(1)$ transformation does not act on this field and the system is still gauge invariant in the $V_\psi$ chart. The transition functions break the semi-local $\mathbb{U}(1)\times\mathbb{S U}(2)$ invariance, leaving us with only a semi-local $\mathbb{U}(1)_{\rm local}\times\mathbb{U}(1)_{\rm global}$ symmetry. The vacuum orbit in this $V_\psi$ chart, the set of zeroes of $U_\psi$, $\psi_1=w_1$,  $\psi_2=w_2$, is the $\mathbb{H}^3$ one-sheet hyperboloid:
\beq
 |w_1|^2-\frac{4\rho^2-a^2}{a^2}\cdot |w_2|^2=\rho^2\cdot \frac{4\rho^2-a^2}{a^2} \quad . \label{hypcp2}
\eeq
Note that even though the hyperboloid (\ref{hypcp2})
is an open space in $\mathbb{C}^2$ it is accommodated within $\mathbb{C P}^2$ through the
\lq\lq infinite line \rq\rq: $\mathbb{C P}^2=\mathbb{C}^2 \cup \mathbb{C P}^1$. We stress, however, that only points complying with (\ref{hypcp2}) and living in $V_\psi$, i.e., $\vert w_1\vert < +\infty,\vert w_2\vert < +\infty$, are bona fide vacua of the system. Like $\mathbb{S}^3$, $\mathbb{H}^3$ is also a 3D fibered space with the $\mathbb{S}^1$-circle
as fibre. The base, however, is one sheet in the 2D hyperboloid living in the $\mathbb{R}^3$ subspace of $\mathbb{C}^2$ where ${\rm Im}\, w_1=0$.
Despite the differences in charges of the fields and vacuum orbit geometry, the Higgs mechanism gives rise to a spectrum formed by a couple
of massive particles, the Higgs and vector bosons, and a complex (but neutral) Goldstone boson. By choosing the vacuum, e.g., on the Equatorial circle of $\mathbb{H}^3$, $(w_1= \rho\cdot \frac{\sqrt{4\rho^2-a^2}}{a},w_2=0)$, we find the same mass spectrum as in the $V_\phi$ chart:
\[
m_A^2=m_H^2=\frac{e^2}{4} a^2 \Big(1-\frac{a^2}{4\rho^2}\Big)^2=m^2 \quad , \quad m_G^2=0 \, \, .
\]
We remark that, despite being \lq\lq neutral\rq\rq, the Goldstone field $\psi_2$ is coupled to the $A_\mu$ vector boson through the mixed terms
induced by the Fubini-Study metric, as can be checked from perturbiations around the vacuum chosen.

The Bogomolny splitting, worked according to the standard procedure, confirms that the Lagrangian  (\ref{lag2}) is again at the self-dual point. The Bogomolny bound is in this $V_\psi$ chart: $E\geq \frac{1}{2} e (4\rho^2-a^2) |\Phi_M|$ where, of course, $4\rho^2$ is greater that $a^2$. Note that it is the same bound as in the north chart of the $\mathbb{S}^2$-sigma model. The Bogomolny equations guaranteeing that this bound is saturated are:
\beq
B=\pm\frac{e}{2}\left(4\rho^2-a^2-\frac{4\rho^2|\psi_1|^2}{\rho^2+|\psi_1|^2+ |\psi_2|^2}\right)\hspace{1cm} D_1\psi_p\pm i D_2\psi_p=0, p=1,2 \label{bog2n}\quad .
\eeq
In particular, the equation between the covariant derivatives of $\psi_2$ in (\ref{bog2n}) is the Cauchy-Riemann equation: the self-dual $\psi_2$
field solutions are holomorphic (+ sign) or antiholomorphic ($-$ sign) functions. The energy per unit length finiteness requires, for configurations reaching a bona fide vacuum $\vert w_2\vert <+\infty$ at infinity, a constant $\psi_2$ value at long distances. Not allowing for the presence of poles, the Liouville theorem, ensuring that an entire and bounded function in $\mathbb{C}$ is constant, requires that $\psi_2$ take a constant value all along the plane. Once this constant value is substituted in the equation for the magnetic field, (\ref{bog2n}) reduces to a system of two equations with the same structure as those giving the self-dual vortices on the north chart of the sphere in Section 4. It then follows that, with very slight modifications, the $\mathbb{S}^2$ solutions found in that section can be embedded in the $V_\psi$ chart, becoming self-dual vortices of the non-linear $\mathbb{C P}^2$ sigma model of the second species. The situation is reminiscent of that studied on the sphere in that regular vortices on $V_\psi$, with a zero of $\psi_1$ at their cores, are singular solutions on the  chart $V_\phi$, whereas regular semi-local vortices on $V_\phi$, where $\phi_1$ is nonzero at infinity, are singular on the chart $V_\psi$. In any case, the vortices on $V_\psi$ are stable solutions, the reason being that the $\psi_1$ field winds at infinity around a two-dimensional throat of $\mathbb{H}^3$ whose radius $\sqrt{(4\rho^2-a^2)(\rho^2+|w_2|^2)}$ is positive regardless of the value of $|w_2|$. Specifically, the choice $\psi_2=0$ leads to exactly the same vortices as those existing in the north chart in the $\mathbb{S}^2$-model. Other constant values of $\psi_2$ merely give rise to a redefinition of the parameters in the solutions.


\subsection{Final comments}

The transition between the charts $V_\phi$ and $V_\xi$ works similarly, and leads to couplings $D_\mu\xi_p=\partial_\mu\xi_p-i N_p e A_\mu\xi_p$ with $N_1=0, N_2=-1$ and a self-dual Lagrangian ${\cal L}_\xi$, which adopts exactly the form (\ref{lag2}), (\ref{pot2}) with the substitutions $\psi_1\rightarrow \xi_2,\psi_2\rightarrow \xi_1$. The self-dual vorticial solutions in this third chart therefore belong to the same species as those found in the $V_\psi$ chart.

The results described along this paper about the existence and structure of self-dual topological defects in the massive gauged $\mathbb{CP}^2$-sigma model suggest a general picture which should be valid for other higher-rank gauged $\mathbb{CP}^n$-sigma non-linear systems. In the $\mathbb{CP}^n$ theory there are $n$ scalar fields and $n+1$ charts, let us call them $V_0,V_1,\ldots,V_n$. Starting with the same coupling for all the fields on the chart $V_0$, we can construct on this chart a self-dual model with semi-local $\mathbb{SU}(n)_{\rm global}\times\mathbb{U}(1)_{\rm local}$ symmetry group, whereas in the other $n$ charts the transition functions would break the global symmetry to a $\mathbb{SU}(n-1)_{\rm global}$ subgroup acting on the neutral fields, and only one field would couple to the $\mathbb{U}(1)_{\rm gauge}$ group through the covariant derivative. There are thus semi-local topological defects on the chart $V_0$ and one-complex-component vortices analogous to those found on the sphere on the remaining ones. The vacuum orbit is $\mathbb{S}^{2n-1}$ for the fields on $V_0$ and $\mathbb{H}^{2n-1}$ for $V_1, V_2,\ldots,V_n$. The stability of the solutions is guaranteed, in the case of $V_0$ by the known arguments for semi-local vortices, and in the other charts because the vorticity of the charged field is due to its asymptotic winding  around the throat of  $\mathbb{H}^{2n-1}$. Some modifications of this scenario could be considered for cases in which the assignation of  couplings to the fields on $V_0$ is different. Let us assume, for instance, that there are on $V_0$ $r$ fields with coupling $N_r e$ and $n-r$ fields with coupling $N_{n-r} e$. The symmetry group in this chart is $\mathbb{SU}(r)_{global}\times \mathbb{SU}(n-r)_{global}\times \mathbb{U}(1)_{gauge}$, and there are two types of semi-local vortices corresponding to the two global factors. All the other charts fit into two different classes. There are $r$ charts such that the transition functions break the symmetry to $\mathbb{SU}(r-1)_{global}\times \mathbb{SU}(n-r)_{global}\times \mathbb{U}(1)_{gauge}$, where the fields complying with the global $\mathbb{SU}(r-1)$ are the neutral ones, those supporting the global $\mathbb{SU}(n-r)$ symmetry have coupling $-(N_r-N_{n-r})e$, and the remaining field has coupling $-N_re$. The solutions existing in these charts are of two kinds: there exist embedded $\mathbb{S}^2$ vortices, where only the field with $-N_r e$ coupling winds, and there are also defects in which this one-complex-component field is mixed with an $\mathbb{SU}(n-r)_{global}\times \mathbb{U}(1)_{gauge}$ semi-local vortex, with winding numbers proportional to the charges; in all these cases, the neutral fields remain frozen  at constant values. In the remaining $n-r$ charts we observe the reciprocal situation, where the symmetry is $\mathbb{SU}(r)_{global}\times \mathbb{SU}(n-r-1)_{global}\times\mathbb{U}(1)_{gauge}$ and the $\mathbb{SU}(n-r-1)$ corresponds to neutral fields, the charged fields have couplings $-(N_{n-r}-N_r)e$ for the global $\mathbb{SU}(r)$, and there is a final charged field which couples to $A_\mu$ with intensity $-N_{n-r} e$.

Finally, we suggest that target manifolds more complicated than ${\mathbb C}{\mathbb P}^N$ could be considered as target manifolds. For instance,
the projective unitary group, the quotient of the ${\mathbb S}{\mathbb U}(N+1)$ group by its center, ${\mathbb P}{\mathbb U}(N+1)=\frac{{\mathbb S}{\mathbb U}(N+1)}{\mathbb{Z}_{N+1}}$, which is a fibered space over ${\mathbb C}{\mathbb P}^N$ of dimension $(N+1)^2-1$ and fiber ${\mathbb U}(N)$, see \cite{boya}, would be a good choice as target space because the possible BPS defects would have
a topology related to the finite group ${\mathbb Z}_{N+1}$ with very interesting properties. In particular, the topology is due to the discrete subgroup $\mathbb{Z}_{N+1}$ a situation of the type discussed in \cite{nitta14} where twisted boundary conditions in the $\mathbb{O}(N)$ sigma model break the discrete symmetry and give rise to fractional vortices,
\section*{Acknowledgement}
After the first version of this work had been completed and sent to the arXives, we learned that vorticial solutions on $\mathbb{S}^2$ like those reported here had been previously found in \cite{nivi} as the partons making two-dimensional $\mathbb{CP}^1$ lumps in a ${\cal N}=(2,2)$ supersymmetric theory. We thank the authors of \cite{nivi} for pointing out  their results to us, which have prompted us to generalize and enlarge the scope of our previous work.

\end{document}